\documentclass[lettersize,journal]{IEEEtran}
\usepackage{amsmath,amssymb,amsfonts}
\usepackage{cite}
\usepackage{algorithmic}
\usepackage{array}
\usepackage{textcomp}
\usepackage{float}
\usepackage{stfloats}
\usepackage{url}
\usepackage{verbatim}
\usepackage{graphicx}
\usepackage{subfigure}
\usepackage{algorithm}
\usepackage{cuted}
\usepackage{stfloats}
\usepackage{multicol} % 控制多栏布局
\usepackage[export]{adjustbox}
\allowdisplaybreaks[4]
\def\BibTeX{{\rm B\kern-.05em{\sc i\kern-.025em b}\kern-.08em
    T\kern-.1667em\lower.7ex\hbox{E}\kern-.125emX}}
\usepackage{balance}

\newtheorem{lemma}{\bf Lemma}

\begin{document}
\bibliographystyle{IEEEtran}
\title{RIS-based Communication Enhancement and Location Privacy Protection in UAV Networks}
\author{Ziqi Chen, \IEEEmembership{Graduate Student Member, IEEE}, Jun Du, \IEEEmembership{Senior Member, IEEE}, Chunxiao Jiang, \IEEEmembership{Fellow, IEEE},\\ Tony Q. S. Quek, \IEEEmembership{Fellow, IEEE} and Zhu Han, \IEEEmembership{Fellow, IEEE}
	\thanks{Z. Chen is with the Department of Electronic Engineering, Tsinghua University, Beijing 100084, China (e-mail: 18801351390@163.com).}
	\thanks{J. Du is with the Department of Electronic Engineering, and also with the State Key Laboratory of Space Network and Communications, Tsinghua University, Beijing 100084, China (e-mail: jundu@tsinghua.edu.cn).}
	\thanks{Chunxiao Jiang is with the Beijing National Research Center for Information Science and Technology, and also with the State Key Laboratory of Space Network and Communications, Tsinghua University, Beijing 100084, China. (email: jchx@tsinghua.edu.cn).}
	\thanks{Tony Q. S. Quek is with the Singapore University of Technology and Design, Singapore 487372, and also with the Yonsei Frontier Lab, Yonsei University, South Korea (e-mail: tonyquek@sutd.edu.sg).}
	\thanks{Z. Han is with the Department of Electrical and Computer Engineering in the University of Houston, Houston, TX 77004 USA, and also with the Department of Computer Science and Engineering, Kyung Hee University, Seoul. South Korea, 446-701 (e-mail: hanzhu22@gmail.com).}
\vspace{-6mm}}

\markboth{TRANSACTIONS ON WIRELESS COMMUNICATIONS, AUGUST~2025}%
{How to Use the IEEEtran \LaTeX \ Templates}

\maketitle

\begin{abstract}
	With the explosive advancement of unmanned aerial vehicles (UAVs), the security of efficient UAV networks has become increasingly critical. Owing to the open nature of its communication environment, illegitimate malicious UAVs (MUs) can infer the position of the source UAV (SU) by analyzing received signals, thus compromising the SU location privacy. To protect the SU location privacy while ensuring efficient communication with legitimate receiving UAVs (RUs), we propose an Active Reconfigurable Intelligent Surface (ARIS)-assisted covert communication scheme based on virtual partitioning and artificial noise (AN). Specifically, we design a novel ARIS architecture integrated with an AN module. This architecture dynamically partitions its reflecting elements into multiple sub-regions: one subset is optimized to enhance the communication rate between the SU and RUs, while the other subset generates AN to interfere with the localization of the SU by MUs. We first derive the Cramér-Rao Lower Bound (CRLB) for the localization with received signal strength (RSS), based on which, we establish a joint optimization framework for communication enhancement and localization interference. Subsequently, we derive and validate the optimal ARIS partitioning and power allocation under average channel conditions. Finally, tailored optimization methods are proposed for the reflection precoding and AN design of the two partitions. Simulation results validate that, compared to baseline schemes, the proposed scheme significantly increases the localization error of the SU by MUs while maintaining efficient communication between the SU and RUs, thereby effectively protecting the SU location privacy.
\end{abstract}

%
%\vspace{-2mm}
\section{Introduction}
In recent years, the burgeoning development of unmanned aerial vehicle (UAV) technology has facilitated its widespread application across an expanding array of emerging domains, including search and rescue, emergency communications, relay communications, and aerial reconnaissance scenarios \cite{zhou2024delay}, \cite{sheng2025towards}, \cite{guo2024joint}. To ensure the proper execution of UAV swarm missions, a reliable, stable, and efficient wireless communication network is essential among UAVs \cite{yu2024secure}. However, the open environment and broadcast nature of wireless communications in UAVs pose significant challenges to the security and reliability of UAV communication networks
\cite{qu2024privacy}, \cite{tong2024uav}. Specifically, during the process of communication among UAVs, illegitimate malicious UAVs (MUs) may exploit the signals they receive to infer the location information of the source UAV (SU). This vulnerability consequently results in the compromise of the SU location privacy, presenting a critical threat to the integrity and confidentiality of UAV operations.

As an emerging paradigm in wireless communications, Reconfigurable Intelligent Surface (RIS) can provide a more stable and reliable communication network for UAV swarms \cite{li2024active} \cite{lin2024multi}. Comprising a large number of controllable reflective units, RIS can dynamically alter the direction, strength, and coverage of electromagnetic waves, thereby enhancing signal quality and improving communication efficiency \cite{truong2025energy}. To protect UAV networks security while enhancing communication, many studies have introduced artificial noise (AN) to achieve privacy protection for UAVs \cite{wen2024ris}, \cite{su2024secure}, \cite{chen2025double}. Specifically, RIS intelligently adjusts its reflection phases to effectively enhance the signal power of the legitimate link, while simultaneously leveraging the AN emitted by the source node, which is reflected by the RIS to generate interference that weakens the illegitimate link \cite{gu2022physical}. However, these studies invariably presuppose that the source node possesses the capability to transmit AN, which constitutes a relatively stringent assumption. Concurrently, the direct joint optimization of communication enhancement and interference at the RIS entails high computational complexity. 

%Many existing studies focus on enhancing communication in underwater environments, but few have considered the issue of location privacy in underwater communication. Due to the openness of the underwater communication environment, the source node (SU) of the transmitted signal is highly susceptible to being located by illegal eavesdropping nodes (MUs) through analysis of the received signals, thereby threatening the privacy and security of the node. In this context, the challenge lies in leveraging Reconfigurable Intelligent Surface (RIS) technology to enhance underwater communication while protecting the location information of the source. Some studies achieve interference with MUs through the synergy of artificial noise (AN) from the transmitter and RIS \cite{han2022artificial}, \cite{arzykulov2023artificial}. However, these studies assume that the source and RIS have prior knowledge of the eavesdropper location, which does not align with real-world eavesdropping scenarios.

To address the above challenges, we innovatively introduce a controllable AN generation module into the existing active RIS (ARIS) architecture, enabling proactive localization interference against MUs. This architecture eliminates the need for the SU to design and transmit AN, thereby enhancing the scalability of the system. In addition, to reduce the complexity of joint optimization, we propose a virtual partitioning mechanism for ARIS, which decouples the multi-objective optimization problem. Specifically, the main contributions of this paper are summarized as follows:
\begin{itemize}
	\item We propose a novel ARIS architecture integrated with an AN module. This innovation enables the ARIS to disrupt the localization attempts of MUs targeting the SU, thereby protecting the SU location privacy. The introduction of the AN module provides architectural support for the multi-objective optimization of communication enhancement and location privacy protection.
%	In contrast to prior studies that predominantly rely on passive RIS or active RIS lacking explicit privacy-preserving mechanisms, this architecture incorporates a controllable AN generator to proactively generate interference. 
	\item We derive the Cramér-Rao Lower Bound (CRLB) for RSS-based localization in an ARIS-assisted communication scenario. This analytical derivation quantifies the fundamental limits of localization accuracy under the influence of AN, providing a benchmark for evaluating the effectiveness of location privacy protection. Meanwhile, we formulate a multi-objective joint optimization problem aimed at simultaneously maximizing the communication rate of legitimate receiving UAVs (RUs) while maximizing the localization error of MUs. This framework effectively couples communication performance with privacy preservation, offering a holistic approach.
	\item We design an innovative virtual partitioning mechanism within the ARIS framework, dynamically dividing the reflecting elements into two functionally distinct subsets: one subset focuses on optimizing the sum rate for RUs, while the other generates AN to interfere with the localization efforts of MUs. This decoupling in the physical domain reduces computational complexity and enhances adaptability. Under average channel conditions, we derive and validate the optimal ARIS partitioning and power allocation, ensuring an effective balance between communication enhancement and localization interference.
	\item Based on the proposed ARIS architecture, we develop tailored optimization algorithms for the two partitioned subsets. For the ARIS partition for communication enhancement (ARIS-CE), we devise an alternating optimization scheme based on Fractional Programming (FP), jointly optimizing the SU beamforming vector and ARIS-CE reflection precoding to achieve near-optimal sum rates. For the ARIS partition for localization interference (ARIS-LI), we design an alternating optimization scheme leveraging FP and Semi-Definite Programming (SDP) to optimize the reflection precoding and AN factors, maximizing the interference-to-signal ratio (ISR) at MUs. Extensive simulations ultimately validate that the proposed approach outperforms baseline methods in communication rate and location privacy protection.
\end{itemize}

The structure of this paper is as follows: Section \ref{work} introduces the related works. Section \ref{sys} presents a detailed exposition of an ARIS-assisted UAV communication model. Section \ref{SBL} describes an RSS-based localization model for MUs and conducts a CRLB analysis. Section \ref{opt} designs a joint optimization scheme for communication enhancement and localization interference. Section \ref{sim} validates the performance and relevant metrics of the proposed mechanism. Finally, Section \ref{concl} is the conclusion of the entire paper. 

\emph{Notations:} $\mathbb{C}$, $\mathbb{R}$ and $\mathbb{R}_+$ denote the sets of complex, real and positive real numbers, respectively. $\left[\cdot\right]^\top$, $\left[\cdot\right]^*$, $\left[\cdot\right]^\dagger$ and $\left[\cdot\right]^{-1}$ represent the transpose, conjugate, conjugate-transpose and inverse operations of a matrix, respectively. $\|\cdot\|$ and $\|\cdot\|_F$ denote the Euclidean norm and the Frobenius norm of the argument, respectively. $\mathbf{I}_K$ is the $K\times K$ identity matrix. $\mathbf{e}_k$ denotes the $k$-th column of the identity matrix. $\mathcal{R}(\cdot)$ and $\mathcal{I}(\cdot)$ denote the real and imaginary parts of the complex-valued arguments, respectively. $\text{Diag}(\cdot)$ forms an $K\times K$ diagonal matrix from a $K$-dimensional vector argument. $\mathcal{CN}\left(\mathbf{\mu}, \mathbf{\Sigma}\right)$ denotes the complex multivariate Gaussian distribution with mean $\mathbf{\mu}$ and variance $\mathbf{\Sigma}$.
%\vspace{-4mm}
\section{Related Works}
\label{work}
\subsection{RIS-assisted Covert Communication with AN}
To provide a stable, reliable, and covert communication network for UAV swarms, many studies have combined AN and RIS to enhance communication for legitimate UAVs while interfering with illegitimate UAVs. Wang $et~al.$ \cite{wang2024active} proposed an AN-based ARIS-assisted covert communication framework that integrates sensing and positioning to counter illegitimate mobile nodes, significantly enhancing interference capabilities against such nodes through joint optimization of power allocation, phase shifts, and UAV positioning. Elsayed $et~al.$ \cite{elsayed2025sum} proposed a joint optimization framework based on Riemannian manifolds, which jointly designs the source beamforming, RIS phase shifts, and AN covariance matrix to enhance the privacy protection effectively. By introducing manifold optimization, the approach significantly reduces computational overhead while satisfying both sensing and communication performance constraints. Wen $et~al.$ \cite{wen2024ris} proposed a RIS-assisted UAV communication scheme against multiple colluding eavesdroppers, jointly optimizing trajectory, beamforming, and RIS phases. Their method pioneers the integration of artificial-noise-aware trajectory design into RIS-UAV systems for enhanced privacy protection. Han $et~al.$ \cite{han2022artificial} proposed an artificial-noise-aided secrecy optimization framework. By jointly optimizing active and passive beamforming, and AN design, the proposed method significantly improves secrecy rates while reducing the required AN power.

Although the introduction of AN can effectively protect the privacy and security of UAV networks, directly solving the multi-objective joint optimization problem at the RIS entails high computational complexity. Moreover, existing studies require the SU to have the capability to transmit AN, which compromises the scalability of the framework. Therefore, it is necessary to introduce a novel ARIS framework capable of emitting AN, along with appropriate mechanism designs, to reduce the complexity of joint optimization.
\vspace{-3mm}
\subsection{RIS with Virtual Partition}
To further improve optimization efficiency, many studies have adopted virtual partitioning mechanisms to decouple the joint optimization problem. Arzykulov $et~al.$ \cite{arzykulov2024aerial} proposed an aerial RIS-aided physical layer security framework that jointly optimizes RIS deployment and virtual partitioning, enabling a single RIS to enhance legitimate communication and AN jamming simultaneously. By deriving closed-form secrecy capacity expressions, the work offers a low-complexity yet highly effective solution for secure wireless communications. Saif $et~al.$ \cite{saif2025ris} proposed a resilient connectivity framework for uplink RIS-assisted UAV networks, introducing a joint optimization of RIS placement and virtual partitioning to maximize algebraic connectivity. This work offers a scalable and robust design for future RIS-enabled aerial-ground integrated networks. Cai $et~al.$ \cite{cai2023ris} proposed a RIS partitioning-based beamforming framework. This work characterizes the fundamental tradeoff between performance and complexity, offering a scalable solution with provable convergence and near-optimal rate in asymptotic and finite-size regimes.

Existing studies on RIS partitioning have primarily focused on passive RIS, which suffers from multiplicative fading in practical applications. Meanwhile, introducing an AN generation module at the RIS necessitates using ARIS to supply the required transmission power for AN. Therefore, beyond conventional RIS partitioning, it is essential to perform judicious power allocation for ARIS further.
\begin{figure}[!t]
	\centering
	\includegraphics[width=0.5\textwidth]{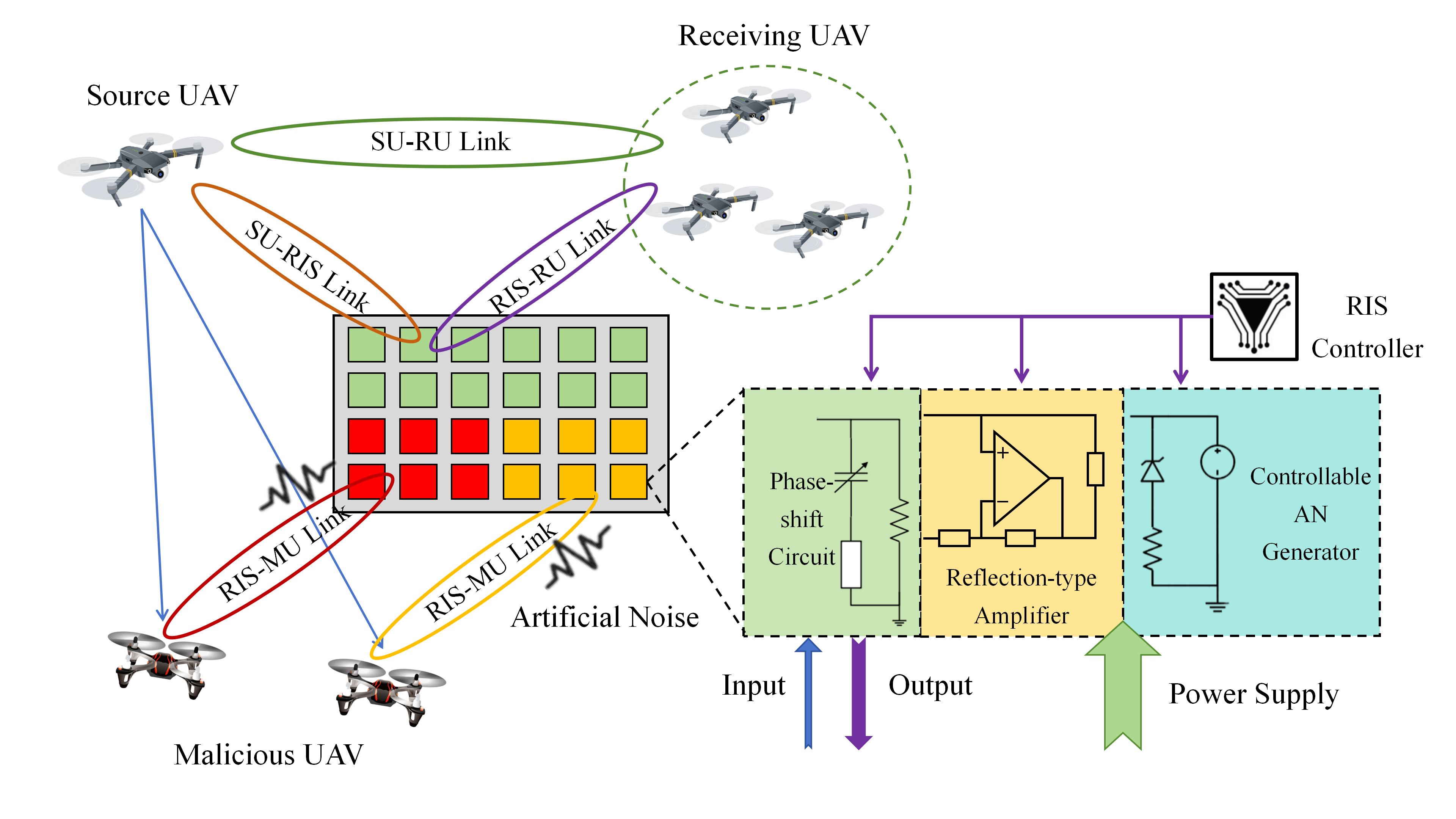}
	\caption{ARIS-assisted UAV networks based on AN and partitioning.}
	\label{FRAMEWORK}
	\vspace{-4mm}
\end{figure}

\section{System Model}
\label{sys}
In this paper, we consider a typical ARIS-assisted UAV communication scenario, which is shown in Fig. \ref{FRAMEWORK}, consisting of one $M$-antenna SU, one ARIS with $N_t$ reflection elements, and $K$ legitimate single-antenna RUs, of which the set is $\mathcal{K} = \left\{1,2,\cdots,K\right\}$. The position coordinates of these three entities are defined as $\mathbf{l}_S\triangleq\left[x_S\ y_S \ z_S\right]^\top$, $\mathbf{l}_{R}\triangleq\left[x_{R}\ y_{R} \ z_{R}\right]^\top$ and $\left\{\mathbf{l}_{r,k} \triangleq\left[x_{r,k}\ y_{r,k} \ z_{r,k}\right]^\top, k\in\mathcal{K}\right\}$, respectively, and assume that the three entities can share location information. Additionally, in the open wireless communication environment, there are $E$ MUs that attempt to illegally locate the SU by analyzing the received signals. The set of MUs is denoted as $\mathcal{E}=\left\{1,2,\cdots,E\right\}$, and their position are defined as $\left\{\mathbf{l}_e\triangleq\left[x_e\ y_e \ z_e\right]^\top, e \in\mathcal{E}\right\}$. We assume that the UAV communication scenario within each time slot is quasi-static, where the SU is static and located sufficiently far from the other UAVs, allowing for a plane wavefront approximation. In addition, we consider MUs can share all information, including the received signals and their respective location.
\vspace{-4mm}
\subsection{Active RIS with Virtual Partition}
Utilizing passive RIS for localization interference depends on externally provided AN, which lacks the ability to control the noise power, and it exhibits limited flexibility and scalability in the adjustment of the AN. Therefore, to enhance communication and protect location privacy of the SU, we introduce an AN module at the RIS to interfere with MUs. The specific framework of the proposed ARIS with an AN module is illustrated in Fig. \ref{FRAMEWORK}. However, jointly optimizing ARIS for communication enhancement and localization interference poses a challenging problem, as it involves complex coupling relationships between the optimization objectives and extremely high computational complexity. To address this, we propose an optimization method based on virtual partitioning, which decouples multi-objective optimization problems by dividing the ARIS reflecting elements into multiple virtual sub-regions. The specific partitioning strategy can be dynamically adjusted according to task requirements: one part of the sub-regions focuses on improving the communication rate for RUs, while the other part generates AN to interfere with signals, explicitly targeting the localization of MUs. Let $\boldsymbol{\rho}=[\rho_0, \rho_1, \cdots,\rho_E]$ and $\boldsymbol{\eta}=[\eta_0,\eta_1,\cdots,\eta_E]$ represent the reflecting elements and the power allocation of the ARIS virtual partition, respectively, which satisfy
\begin{equation}
	\rho_0 + \sum_{e=1}^{E}\rho_e= \eta_0 + \sum_{e=1}^{E}\eta_e=1,
\end{equation}
where $\rho_0$ and $\eta_0$ represent the proportion of ARIS reflecting elements and the power allocated for ARIS-CE, respectively, while $\rho_e$ and $\eta_e$ represent the proportion of ARIS reflecting elements and the power allocated for 
ARIS-LI of MU $e$, respectively. Meanwhile, we define $\mathcal{N}_0 = \left\{1,2,\cdots,N_0\right\}$ and $\mathcal{N}_e = \left\{1,2,\cdots,N_e\right\}$ as the set of ARIS-CE and ARIS-LI, respectively. Ultimately, the number of ARIS reflection elements used for communication enhancement and the number of ARIS reflection elements used for interfering with the
localization of MU $e$ can be expressed as
%\vspace{-1mm}
\begin{equation}
	N_0 = \lfloor \rho_0 N_t \rfloor\ , N_e = \lfloor \rho_e N_t \rfloor\ \!\!.
\end{equation}

According to the proposed ARIS architecture, the reflected and amplified signals of the ARIS-CE is expressed as
\begin{subequations}
	\begin{align}
		&\mathbf{y}_0 = \mathbf{\Theta} \mathbf{x} + \mathbf{\Theta} \mathbf{v}_0 + \mathbf{n}_0,\ \mathbf{\Theta}=\text{Diag}(\boldsymbol{\theta^\top}),\label{1a}\\
		&\boldsymbol{\theta} = \left[p_1 e^{j\theta_1},p_2 e^{j\theta_2},\cdots,p_{N_0} e^{j\theta_{N_0}}\right]^\top\in \mathbb{C}^{{N_0}\times 1},\vspace{-3mm} \label{1b}
	\end{align}
\end{subequations}
where $\mathbf{\Theta}$ represents the reflection precoding of the ARIS-CE, in which $p_n \in \mathbb{R}_{+}$ and $\theta_n$ denote the amplification factor and the phase shift factor of element $n\in\mathcal{N}$. Meanwhile, the active characteristics of ARIS introduce corresponding noise, which can be categorized into dynamic noise $\mathbf{\Theta} \mathbf{v}_0$ and static noise $\mathbf{n}_0$. Specifically, $\mathbf{v}_0$ is the intrinsic noise (IN), including the input noise of ARIS and the inherent device noise, with $\mathbf{v}_0\sim\mathcal{CN}(\mathbf{0}_{N_0}, \sigma_v^2\mathbf{I}_{N_0})$, where $\sigma_v^2$ is a fixed constant.  $\mathbf{n}_0$ mainly represents the noise generated by the phase shift circuit. Since the energy of $\mathbf{n}_0$ is extremely small compared to $\mathbf{\Theta}\mathbf{v}_0$, $\mathbf{n}_0$ is neglected in the subsequent analysis. 

Meanwhile, the reflected signals of the ARIS-LI can be expressed as
\begin{subequations}
	\begin{align}
		&\mathbf{y}_e = \mathbf{\Theta}_e \mathbf{x} + \mathbf{\Theta}_e \mathbf{v}_e + \mathbf{n}_0,\ \mathbf{\Theta}_e=\text{Diag}(\boldsymbol{\theta}^\top_e),\label{2a}\\
		&\boldsymbol{\theta}_e = \left[e^{j\phi_1}, e^{j\phi_2},\cdots,e^{j\phi_{N_e}}\right]^\top\in \mathbb{C}^{{N_e}\times 1},\vspace{-3mm} \label{2b}\vspace{-2mm}
	\end{align}
\end{subequations}
where $\mathbf{\Theta}_e$ represents the reflection precoding of the ARIS-LI. To achieve effective interference with MUs, the RIS partition adopts a reflection mechanism similar to that of a passive RIS, utilizing the adjustment of the reflecting phases along with the transmission of AN to perform localization interference. Thus, $\phi_n$ denotes the phase shift factor of element $n\in\mathcal{N}_e$. Additionally, the noise generated at the ARIS can be considered as the AN to interfere with the MUs \cite{lyu2023robust}. Therefore, to further protect the location privacy of the SU, we introduce a controllable noise generator to dynamically adjust AN $\mathbf{v}_e$ introduced at the ARIS. 

%Specifically, based on the structure shown in Fig. \ref{UARIS}, we can model the AN $\mathbf{v}$ as $\mathbf{v}\sim\mathcal{CN}(\mathbf{0}_{N_e}, \sigma_{e}^2\mathbf{I}_{N_e})$, wherein $\sigma_{e}^2$ denotes the adjustable AN power generated at the ARIS.
%\begin{figure}[!t]
%	\centering
%	\includegraphics[width=3.2in]{UARIS.png}
%	\caption{The hardware architectures of a UARIS with a controllable noise generator.}
%	\label{UARIS2}
%\end{figure}

\subsection{Channel Model}
To investigate the relationship between distances and path losses, without loss of generality, the far-field spherical-wave propagation model is applied to characterize the large-scale fading of the channel, which is formulated as
\begin{equation}
	L=L_0d^{-\epsilon }\vspace{-2mm},
\end{equation}
%According to the 3GPP standard, the path loss model to characterize the large-scale fading of the channels can be formulated as 
%\begin{equation}
%	L = 37.3 + 22.0 \lg\left(d\right),\label{pl}
%\end{equation}
where $d$ is the distance between two nodes and $L_0$ is the path loss at the reference distance of 1 m. Additionally, $\epsilon$ denotes the path loss exponent. Furthermore, to characterize the small-scale fading, we utilize the Rayleigh fading channel model for all channels involved. Therefore, an arbitrary channel gain between two nodes can be modeled as
\begin{equation}\label{G}
	F = \sqrt{L}\Gamma, \Gamma\sim\mathcal{CN}\left(0, \sigma_F^2\right),
\end{equation}
where $\sigma_F^2$ represents the variance of the Rayleigh fading distribution, and we assume $\sigma_F^2 = 1$ in this paper. To facilitate the subsequent derivation of relevant conclusions under the average channel, we introduce the following lemma.
\begin{lemma}\label{l4}
		Assuming that $f$ is the channel gain generated by (\ref{G}), the following expectation can be derived:
		\begin{equation}
				\mathbb{E} \left[|f|\right]=\frac{\sqrt{\pi}L_f}{2},\mathbb{E} \left[|f|^2\right]=L_f,
			\end{equation}
		where $L_f$ is the path loss of $f$.
		 
		$\mathit{Proof.}$ Detailed proof can be found in \cite{sklar1997rayleigh}.
\end{lemma}

 As described in \cite{hu2021two} and \cite{abdallah2022ris}, various methods can be used to obtain the CSI between SU, RUs and the ARIS. For the purpose of our analysis, we consider that MUs have access to perfect CSI since they are active users within the network, as discussed in \cite{mamaghani2022aerial}. Therefore, considering global CSI, SU can exchange information with the ARIS controller through a dedicated wireless channel \cite{wu2019towards}, jointly optimizing the beamforming vector of the source node and the reflection matrix of the ARIS-CE. Furthermore, the RIS controller can also optimize the reflection matrix of the ARIS-LI based on the CSI information between MUs and the ARIS.

\subsection{ARIS-assisted Communication Model}\label{com}
In this subsection, we will provide a detailed introduction to the ARIS-assisted communication model. Specifically, we consider a downlink UAV communication scenario and denote the symbol vector transmitted to $K$ RUs as $\mathbf{q}:=\left[q_1, q_2, \cdots, q_K\right]^\top \in \mathbb{C}^{K\times 1}$, which satisfies $\mathbb{E}\left\{\mathbf{q}\mathbf{q}^\dagger\right\}=\mathbf{I}_K$. As shown in Fig. \ref{FRAMEWORK}
, let $\mathbf{h}_k\in \mathbb{C}^{M\times 1}$, $\mathbf{H}\in \mathbb{C}^{N_0\times M}$, $\mathbf{H}_e\in \mathbb{C}^{N_e\times M}$, $\mathbf{g}_{k, 0}\in \mathbb{C}^{N_0\times 1}$ and $\mathbf{g}_{k, e}\in \mathbb{C}^{N_e\times 1}$ denote the channel matrix between the SU and RU $k$, that between the SU and the ARIS-CE, that between the SU and ARIS-LI $e$, that between the ARIS-CE and RU $k$, and that between ARIS-LI $e$ and RU $k$. Then, according to (\ref{1a}), signal $r_k\in\mathbb{C}$ received at RU $k$ can be modeled as
\begin{equation}
	\begin{split}
		r_k = \underbrace{\mathbf{h}_k^\dagger\mathbf{s}}_{\textbf{aligned data}} + &\underbrace{\mathbf{g}_{k,0}^\dagger \mathbf{\Theta} \mathbf{H}\mathbf{s} + \sum_{e=1}^{E}\mathbf{g}_{k,e}^\dagger \mathbf{\Theta}_e \mathbf{H}_e\mathbf{s}}_{\textbf{non-aligned data}}\\
		&+\underbrace{\mathbf{g}_{k,0}^\dagger \mathbf{\Theta} \mathbf{v}_0}_{\textbf{non-aligned IN}} +\underbrace{\sum_{e=1}^{E}\mathbf{g}_{k,e}^\dagger \mathbf{\Theta}_e \mathbf{v}_e}_{\textbf{non-aligned AN}} + n_k,\label{4}
	\end{split}
\end{equation}
where $\mathbf{s} = \sum_{k=1}^{K} \mathbf{w}_k q_k$, and $\mathbf{w}_k\in\mathbb{C}^{M\times 1}$ denotes the beamforming vector designed by the SU for symbol $q_k$. Additionally, $\mathbf{v}_e$ denotes the AN introduced by the ARIS, while $n_k$ denotes the background noise introduced at RU $k$, with $n_k\sim \mathcal{CN}(0, \sigma^2)$. Similarly, the received signal model at MU $e$ can be expressed as
\begin{equation}
	\begin{split}
		r_e = \underbrace{\mathbf{h}_e^\dagger\mathbf{s}}_{\textbf{aligned data}} + &\underbrace{\mathbf{g}_{e,0}^\dagger \mathbf{\Theta} \mathbf{H}\mathbf{s} + \sum_{i=1}^{E} \mathbf{g}_{e,i}^\dagger\mathbf{\Theta}_i \mathbf{H}_i\mathbf{s}}_{\textbf{non-aligned data}}\\
		&+\underbrace{\mathbf{g}_{e,0}^\dagger \mathbf{\Theta} \mathbf{v}_0}_{\textbf{non-aligned IN}} +\underbrace{\sum_{i=1}^{E} \mathbf{g}_{e,i}^\dagger\mathbf{\Theta}_i \mathbf{v}_i}_{\textbf{AN}} + n_e,\label{5}
	\end{split}
\end{equation}
where $\mathbf{h}_e\in \mathbb{C}^{M\times 1}$, $\mathbf{g}_{e, 0}\in \mathbb{C}^{N_0\times 1}$ and $\mathbf{g}_{e, i}\in \mathbb{C}^{N_i\times 1}$ denote the channel matrix between the SU and MU $e$, that between the ARIS-CE and MU $e$, and that between ARIS-LI $i$ and MU $e$. Additionally, $n_e$ denote the background noise introduced at MU $e$, with $n_e\sim \mathcal{CN}(0, \sigma^2)$.

ARIS broadcasts the pre-shared CSI and the generated AN to each RU through a dedicated channel, enabling them to filter out the AN from the received signals effectively \cite{niu2022active}. Furthermore, with the optimization of the ARIS reflection precoding, we consider that the reflective elements in the ARIS-CE are fully aligned with the cascaded SU-RUs channel. Meanwhile, the reflective elements in the ARIS-LI are aligned with the SU-MUs channel but misaligned with the cascaded SU-RUs channel. The optimization method of the ARIS reflection precoding will be discussed in Section \ref{opt}. To facilitate the analysis, we ignore the impact of misaligned channels on the RUs \cite{arzykulov2024aerial}. Therefore, the signal-to-interference-plus-noise ratio (SINR) at RU $k$ is expressed as
\begin{subequations}
	\begin{align}
		&\gamma_k = \frac{|\overline{\mathbf{h}}_k^\dagger \mathbf{w}_k|^2}{\sum_{a=1, a \neq k}^{K} |\overline{\mathbf{h}}_k^\dagger \mathbf{w}_a|^2 + \left\| \mathbf{g}_{k,0}^\dagger \mathbf{\Theta} \right\|^2 \sigma_v^2 + \sigma^2},\\
		&\overline{\mathbf{h}}_k^\dagger=\mathbf{h}_k^\dagger + \mathbf{g}_{k,0}^\dagger \mathbf{\Theta} \mathbf{H}=\mathbf{h}_k^\dagger +\boldsymbol{\theta}^\top\text{Diag}(\mathbf{g}_{k,0}^\dagger)\mathbf{H}
%		\in\mathbb{C}^{1\times M}
	\end{align}
\end{subequations}

However, the MUs do not have prior knowledge of the AN and can only treat it as interference. Meanwhile, since subpartition $e$ of ARIS-LI is only aligned with its corresponding SU-MU $e$ cascaded channels, we similarly ignore the interference of AN from misaligned channels on MU $e$ for
tractability reasons \cite{arzykulov2024aerial}. Furthermore, we consider an interference-limited scenario for MUs, where the power of additive white Gaussian noise and the IN of the ARIS-CE can be neglected due to their significantly smaller magnitude compared to the interference. Thus, the ISR at MU $e$ can be expressed as
\begin{subequations}
	\begin{align}
		&\kappa_e= \frac{\left| \mathbf{g}_{e,e}^\dagger \mathbf{\Theta}_e \mathbf{v}_e\right|^2}{\sum_{k=1}^{K}\left|\left(\overline{\mathbf{h}}_e^\dagger + \mathbf{g}_{e,e}^\dagger \mathbf{\Theta}_e \mathbf{H}_e \right)\mathbf{w}_k\right|^2},\\
		&\overline{\mathbf{h}}_e^\dagger=\mathbf{h}_e^\dagger + \mathbf{g}_{e,0}^\dagger \mathbf{\Theta} \mathbf{H}.
	\end{align}
\end{subequations}
%\begin{figure}[!t]
%	\centering
%	\includegraphics[width=3.2in]{DOWNLINK.png}
%	\caption{UARIS-assisted underwater acoustic downlink communication system.}
%	\label{DOWNLINK}
%\end{figure}
%
%According to the communication model in \ref{4}, the signal-to-interference-plus-noise ratio (SINR) at RU $k$ can be formulated as
\section{RSS-based Localization Method of the Malicious Nodes}\label{SBL}
In this section, we provide a detailed introduction to the RSS model in the ARIS-assisted communication scenario and present the corresponding localization algorithm. Additionally, we theoretically derive the CRLB of the localization model, providing a theoretical foundation for subsequent optimization problem modeling.
\subsection{RSS Model of the Malicious Nodes}
According to the analysis in Section \ref{com}, the relevant signal components that primarily affect the RSS measurements at MU e can be reformulated as
\begin{equation}
	\begin{split}
		\overline{r}_e = \underbrace{\mathbf{h}_e^\dagger\mathbf{s}}_{\textbf{Direct Path}} + \underbrace{\mathbf{g}_{e,0}^\dagger \mathbf{\Theta} \mathbf{H\mathbf{s}} + \mathbf{g}_{e,e}^\dagger \mathbf{\Theta}_e \mathbf{H}_e\mathbf{s}}_{\textbf{Reflected Path}}+\underbrace{\mathbf{g}_{e,e}^\dagger \mathbf{\Theta}_e \mathbf{v}_e}_{\textbf{Noise}}.
	\end{split}
\end{equation}

%\begin{equation}
%	\begin{split}
%		\overline{r}_e = \underbrace{\mathbf{g}_{e,0}^\dagger \mathbf{\Theta} \mathbf{H\mathbf{W}} + \mathbf{g}_{e,e}^\dagger \mathbf{\Theta}_e \mathbf{H}_e\mathbf{W}}_{\textbf{Reflected Path}},
%	\end{split}
%\end{equation}
%where $\mathbf{g}_{e, 0}\in \mathbb{C}^{N_0\times 1}$ and $\mathbf{g}_{e, i}\in \mathbb{C}^{N_i\times 1}$ denote the channel matrix between the SU and MU $e$, that between the ARIS-CE and MU $e$, and that between ARIS-LI $i$ and MU $e$.
%\begin{subequations}
%	\begin{align}
%		&\mathbf{y}_0 = \mathbf{\Theta} \mathbf{x} + \mathbf{\Theta} \mathbf{v}_0 + \mathbf{n}_0,\ \mathbf{\Theta}=\text{Diag}(\boldsymbol{\theta^\top}),\label{1a}\\
%		&\boldsymbol{\theta} = \left[p_1 e^{j\theta_1},p_2 e^{j\theta_2},\cdots,p_{N_0} e^{j\theta_{N_0}}\right]^\top\in \mathbb{C}^{{N_0}\times 1}, \label{1b}
%	\end{align}
%\end{subequations}
%\begin{subequations}
%	\begin{align}
%		&\mathbf{y}_e = \mathbf{\Theta}_e \mathbf{x} + \mathbf{\Theta} \mathbf{v}_e + \mathbf{n}_0,\ \mathbf{\Theta}_e=\text{Diag}(\boldsymbol{\theta}^\top_e),\label{2a}\\
%		&\boldsymbol{\theta}_e = \left[e^{j\phi_1}, e^{j\phi_2},\cdots,e^{j\phi_{N_e}}\right]^\top\in \mathbb{C}^{{N_e}\times 1}, \label{2b}
%	\end{align}
%\end{subequations}

The noise power at MU $e$ can be formulated as 
\begin{equation}
	\begin{split}
		&P_{n,e} = \mathbb{E}\left[\left|\mathbf{g}_{e,e}^\dagger \mathbf{\Theta}_e \mathbf{v}_e\right|^2\right]=P_{v,e}L_{g,e}=P_{v,e}L_0d_{g,e}^{-\epsilon },
	\end{split}
\end{equation}
\begin{equation}
	\begin{split}
		&p_{n,e} = 10\lg P_{n,e}=10\lg P_{v,e} + 10\lg L_0- 10\epsilon\lg d_{g,e},
	\end{split}
\end{equation}
where $P_{v,e} = \left\|\mathbf{v}_e\right\|^2$ denotes the AN power of ARIS-LI $e$ and $d_{g,e}$ represents the distance between MU $e$ and the ARIS.

According to \cite{martin2012modeling}, the signal strength of the direct path is calculated as
\begin{equation}
	\begin{split}
		P_{d,e} &= \mathbb{E}\left[\left|\mathbf{h}_e^\dagger\mathbf{s}\right|^2\right]=\mathbb{E}\left[\mathbf{s}^\dagger\mathbf{h}_e\mathbf{h}_e^\dagger\mathbf{s}\right]\\
		& =\mathbb{E}\left[\text{tr}\left(\mathbf{h}_e\mathbf{h}_e^\dagger\mathbf{s}\mathbf{s}^\dagger\right)\right] = \text{tr}\left(\mathbb{E}\left[\mathbf{h}_e\mathbf{h}_e^\dagger\right]\mathbb{E}\left[\mathbf{s}\mathbf{s}^\dagger\right]\right).
	\end{split}\label{pd}
\end{equation}

According to (\ref{G}), we obtain
\begin{equation}
	\mathbb{E}\left[\mathbf{h}_e\mathbf{h}_e^\dagger\right] = L_{h,e}\mathbf{I}_M=L_0d_{h,e}^{-\epsilon }\mathbf{I}_M, \label{hh}
\end{equation}
where $d_{h,e}$ denotes the distance between SU and MU $e$.

Next, we compute $\mathbb{E}\left[\mathbf{s}\mathbf{s}^\dagger\right]$ as follows:
\begin{equation}
	\begin{split}
		&\mathbb{E}\left[\mathbf{s}\mathbf{s}^\dagger\right]=\sum_{k=1}^{K}\sum_{l=1}^{K}\mathbf{w}_k \mathbb{E}\left[q_k q_l^*\right] \mathbf{w}_l^\dagger\\
		&=\sum_{k=1}^{K}\mathbf{w}_k \mathbb{E}\left[q_k q_k^*\right] \mathbf{w}_k^\dagger = \sum_{k=1}^{K}\mathbf{w}_k \mathbf{w}_k^\dagger.\label{ss}
	\end{split}
\end{equation}

Substituting (\ref{hh}) and (\ref{ss}) into (\ref{pd}), we obtain
\begin{equation}
	\begin{split}
		P_{d,e}=L_{h,e}\text{tr}\left(\sum_{k=1}^{K}\mathbf{w}_k \mathbf{w}_k^\dagger\right)=P_SL_0d_{h,e}^{-\epsilon },
	\end{split}
\end{equation}
\begin{equation}
	\begin{split}
		p_{d,e}=10\lg P_{h,e}=p_S + 10\lg L_0-10\epsilon \lg d_{h,e},
	\end{split}
\end{equation}
where $P_S = \sum_{k=1}^{K}\left\|\mathbf{w}_k\right|^2$ represents the total transmission power of the SU and $p_S = 10\lg P_S$.
% Further, converting $p_d$ into the dB domain, we obtain
%\begin{equation}
%	p_{d, e} = p_{S} - 37.3 - 22\lg d_{h,e},
%\end{equation}
%
%where $p_{S} = 10\lg\left(P_S\right)$ denotes the total transmission power of the SU in the dB domain, and $d_{h,e}$ denotes the distance between the SU and MU $e$.

Similarly, the signal strength of the reflected path is calculated as
\begin{equation}\label{pr}
	\begin{split}
		&P_{r,e} = \mathbb{E}\left[\left|\left(\mathbf{g}_{e,0}^\dagger \mathbf{\Theta} \mathbf{H} + \mathbf{g}_{e,e}^\dagger \mathbf{\Theta}_e \mathbf{H}_e\right)\mathbf{s}\right|^2\right]\\
		&=\mathbb{E}\left[\mathbf{g}_{e,0}^\dagger \mathbf{\Theta} \mathbf{H}\mathbf{s}\mathbf{s}^\dagger\mathbf{H}^\dagger\mathbf{\Theta}^\dagger\mathbf{g}_{e,0}\right] + \mathbb{E}\left[\mathbf{g}_{e,e}^\dagger \mathbf{\Theta}_e \mathbf{H}_e\mathbf{s}\mathbf{s}^\dagger\mathbf{H}_e^\dagger\mathbf{\Theta}_e^\dagger\mathbf{g}_{e,e}\right].
	\end{split}
\end{equation}

First, we solve the first term of (\ref{pr}) and obtain
\begin{equation}
	\begin{split}
		&\mathbb{E}\left[\mathbf{g}_{e,0}^\dagger \mathbf{\Theta} \mathbf{H}\mathbf{s}\mathbf{s}^\dagger\mathbf{H}^\dagger\mathbf{\Theta}^\dagger\mathbf{g}_{e,0}\right] = \mathbb{E}\left[\text{tr}\left(\mathbf{g}_{e,0}^\dagger \mathbf{\Theta} \mathbf{H}\mathbf{s}\mathbf{s}^\dagger\mathbf{H}^\dagger\mathbf{\Theta}^\dagger\mathbf{g}_{e,0}\right)\right] \\
%		& = \mathbb{E}\left[\text{tr}\left(\mathbf{g}_{e,0}^\dagger\mathbf{g}_{e,0} \mathbf{\Theta}\mathbf{\Theta}^\dagger \mathbf{H}\mathbf{s}\mathbf{s}^\dagger\mathbf{H}^\dagger\right)\right]\\
		& = 
		\text{tr}\left(\mathbb{E}\left[\mathbf{g}_{e,0}^\dagger\mathbf{g}_{e,0}\right]\text{Diag}\left(p_1^2,\cdots,p_{N_0}^2\right)\mathbb{E}\left[\mathbf{H}\mathbf{s}\mathbf{s}^\dagger\mathbf{H}^\dagger\right]\right)\\
		& =L_{g,e}\text{tr}\left(\text{Diag}\left(p_1^2,\cdots,p_{N_0}^2\right)\mathbb{E}\left[\mathbf{H}\mathbf{s}\mathbf{s}^\dagger\mathbf{H}^\dagger\right]\right).
	\end{split}\label{pr1}
\end{equation}

Let $\mathbf{A} = \mathbb{E}\left[\mathbf{H}\mathbf{s}\mathbf{s}^\dagger\mathbf{H}^\dagger\right]$, and denote $A_{i,j}$ as the element in the $i$-th row and $j$-th column of matrix $\mathbf{A}$. Then, we obtain
\begin{equation}
	\begin{split}\label{A}
		A_{i,j} &= \sum_{m=1}^M \sum_{n=1}^M \mathbb{E}\left[s_m s_n^*\right] E[H_{i,m} H_{j,n}^*]\\
		&=L_{H,e}\sum_{m=1}^M \sum_{n=1}^M \mathbb{E}\left[s_m s_n^*\right]\delta_{i,j} \delta_{m,n}\\
		&=L_{H,e}\mathbb{E}\left[\mathbf{s}^\dagger\mathbf{s}\right]\delta_{i,j}=L_{H,e}P_S\delta_{i,j},
	\end{split}
\end{equation}
where $s_a$ represents the $a$-th element of $\mathbf{s}$, and $H_{a, b}$ represents the element in the $a$-th row and $b$-th column of matrix $\mathbf{H}$.

Substituting (\ref{A}) into (\ref{pr1}), we obtain
\begin{equation}
	\mathbb{E}\left[\mathbf{g}_{e,0}^\dagger \mathbf{\Theta} \mathbf{H}\mathbf{s}\mathbf{s}^\dagger\mathbf{H}^\dagger\mathbf{\Theta}^\dagger\mathbf{g}_{e,0}\right] = L_{g,e}L_{H,e}P_S\sum_{n=1}^{N_0}p_n^2.
\end{equation}

Similarly, we can derive that the second term of (\ref{pr}) is
\begin{equation}
	\mathbb{E}\left[\mathbf{g}_{e,e}^\dagger \mathbf{\Theta}_e \mathbf{H}_e\mathbf{s}\mathbf{s}^\dagger\mathbf{H}_e^\dagger\mathbf{\Theta}_e^\dagger\mathbf{g}_{e,e}\right] = L_{g,e}L_{H,e}P_SU_e.
\end{equation}

Finally, the signal strength of the reflected path can be expressed as
\begin{equation}
	P_{r,e} = P_S\left(\sum_{n=1}^{N_0}p_n^2 + N_e\right)L_0^2\left(d_{g,e}d_{H}\right)^{-\epsilon},
\end{equation}
\begin{equation}
	p_{r,e} = p_S+20\lg L_0-10\epsilon\lg\left(d_{g,e}d_{H}\right)+G_R,
\end{equation}
where $G_{R} =10\lg\left(\sum_{n=1}^{N_0}p_n^2 + N_e\right)$ is an unknown variable of MUs related to ARIS to be estimated. $d_{H}$ denotes the distance between the ARIS and the SU.

According to the RSS model proposed in \cite[Section II-C]{martin2012modeling}, a reasonable approximation of the RSS measured at MUs is to add means of the signal and noise in the linear domain, which is formulated as
\begin{equation}\label{model}
	\mathbf{R}\sim\mathcal{N}\left(\mathbf{c}, \sigma_{dB}^2\mathbf{I}_E\right),
\end{equation}
\begin{equation}
	c_e = 10\lg\left(10^{p_{d,e}/10} + 10^{p_{r,e}/10} +10^{p_{n,e}/10}
	\right),
\end{equation}
where $\sigma_{dB}^2$ is the variance of the measurement error caused by channel shadow fading.

\subsection{CRLB Analysis of RSS Localization Model}
According to the RSS model in (\ref{model}), we will propose the corresponding localization method and provide a detailed derivation of the CRLB for this localization model. The parameter vector $\mathbf{\Pi}$ to be estimated is expressed as $\mathbf{\Pi} = \left[x_S, y_S, z_S, G_{R}, p_S\right]$. Then, we derive the MLE for the RSS model in (\ref{model}), which is defined as
\begin{subequations}\label{MLE}
	\begin{align}
		\hat{\mathbf{\Pi}}_{\text{ML}} =& \arg\max_{\mathbf{\Pi}}\mathcal{M},\\ \mathcal{M} = -\frac{E}{2}\ln\left(2\pi\sigma_{dB}^2\right)&-\frac{\sum_{e=1}^{E}\left(R_e-c_e\right)}{2\sigma_{dB}^2}.
	\end{align}
\end{subequations}

Due to the normal distribution of the RSS model in (\ref{model}), the MLE in (\ref{MLE}) can be equivalently expressed as
\begin{subequations}\label{lo}
	\begin{align}
		\mathbf{R}&\sim\mathcal{N}\left(\mathbf{c}\left(\mathbf{\Pi}\right), \sigma_{dB}^2\mathbf{I}_E\right),\\
		\hat{\mathbf{\Pi}}_{\text{ML}} &= \arg\min_{\mathbf{\Pi}}\left\|\mathbf{R} - \mathbf{c}\left(\mathbf{\Pi}\right)\right\|.
	\end{align}
\end{subequations}

The localization problem (\ref{lo}) can be solved via the grid search or Newton-Raphson, which will not be elaborated further in this paper. Next, we conduct a CRLB analysis for this RSS localization model. According to (\ref{MLE}), the Fisher Information Matrix (FIM) is given by
\begin{equation}
	\mathbf{J}=\mathbb{E}\left[\left(\nabla_{\mathbf{\Pi}}\mathcal{L}\right)\left(\nabla_{\mathbf{\Pi}}\mathcal{L}\right)^\top\right]=\frac{1}{\sigma_{dB}^2}\sum_{e=1}^{E}\left(\nabla_{\mathbf{\Pi}}c_e\right)\left(\nabla_{\mathbf{\Pi}}c_e\right)^\top.
\end{equation}

Let $\beta_e = 10^{p_{d,e}/10} + 10^{p_{r,e}/10} +10^{p_{n,e}/10}$, then we obtain
\begin{equation}\label{ce}
	\begin{split}
		\nabla_{\mathbf{\Pi}}c_e=\frac{1}{\beta_e}\left(10^{\frac{p_{d,e}}{10}}\nabla_{\mathbf{\Pi}}p_{d,e}+10^{\frac{p_{r,e}}{10}}\nabla_{\mathbf{\Pi}}p_{r,e}\right).
	\end{split}
\end{equation}

Take the first derivative of (\ref{ce}) with respect to the parameter to be estimated and let $e_{dB} = \frac{10\epsilon}{\ln 10}$, then we obtain
\begin{subequations}\label{nb}
	\begin{align}
		\nabla_{\mathbf{\Pi}}p_{d,e}\!\! =\!\! \left[\frac{e_{dB}}{d_{h,e}}\mathbf{u}_{h,e}^\top,0,1\right]^\top\!\!\!,&\nabla_{\mathbf{\Pi}}p_{r,e} \!\!=\!\! \left[\frac{e_{dB}}{d_{H}}\mathbf{u}_{H}^\top,1,1\right]^\top\!\!\!,\\
		\mathbf{u}_{h,e} = \frac{\mathbf{l}_e - \mathbf{l}_S}{d_{h,e}},& \mathbf{u}_{H} = \frac{\mathbf{l}_R - \mathbf{l}_S}{d_{H}}.\label{u}
	\end{align}
\end{subequations}

For the convenience of subsequent derivations, we define the following factors:
\begin{subequations}\label{alp}
	\begin{align}
		&~~\alpha_{0,e} = \frac{10^{\frac{p_{d,e}}{10}}+10^{\frac{p_{r,e}}{10}}}{10^{\frac{p_{d,e}}{10}}+10^{\frac{p_{r,e}}{10}}+10^{\frac{p_{n,e}}{10}}} = \frac{1}{1+\varrho_e},\\
		&~~~~~~~~~~~~~~\alpha_{e} = \frac{10^{\frac{p_{r,e}}{10}}}{10^{\frac{p_{d,e}}{10}}+10^{\frac{p_{r,e}}{10}}},
	\end{align}
\end{subequations}
where $\varrho_e$ represents the expectation of the ISR at MU $e$.

Substituting (\ref{alp}) and (\ref{nb}) into (\ref{ce}), we obtain
\begin{equation}\label{vy}
	\begin{split}
		&\mathbf{v}_e = \left(1-\alpha_{e}\right)\nabla_{\mathbf{\Pi}}p_{d,e} + \alpha_{e}\nabla_{\mathbf{\Pi}}p_{r,e}\\
		& = \left[e_{dB}\left(\frac{1-\alpha_{e}}{d_{h,e}}\mathbf{u}_{h,e} + \frac{\alpha_{e}}{d_{H}}\mathbf{u}_{H}\right)^\top \!\!\!, \alpha_{e}, 1\right]^\top.
	\end{split}
\end{equation}
\begin{equation}\label{crlb}
	\begin{split}
		\sqrt{\text{COV}}\left[\mathbf{l}_{S}\right]&\geq\left(\sqrt{\mathbf{J}^{-1}}\right)_{3\times3}\\
		&=\sigma_{dB}\left(\sqrt{\left(\sum_{e=1}^{E}\frac{\mathbf{v}_{e}\mathbf{v}_{e}^\top}{\left(1+\varrho_e\right)^2}\right)^{-1}}\right)_{3\times3}.
	\end{split}
\end{equation}

By analyzing (\ref{crlb}), Under ideal conditions, the CRLB is proportional to $\varrho_e$, providing a theoretical foundation for the subsequent optimization problem formulation in Section \ref{opt}.
\vspace{-3mm}
\section{Joint Transmit Bemaforming and Reflect Precoding Design with Artificial Noise}\label{opt}
\vspace{-1mm}
To investigate the communication enhancement and location privacy protection supported by the ARIS with partition in UAV communication scenarios, we will model a multi-objective optimization problem for this scenario in this section. Meanwhile, we derive the optimal solution for the ARIS partition and power allocation under the average channel conditions. Furthermore, a joint optimization scheme for transmit beamforming, reflection precoding, and AN is proposed.
\vspace{-4mm}
\subsection{Problem Formulation}
In the UAV communication scenario, we need to enhance communication for RUs while interfering with localization for MUs. Therefore, our optimization problem aims to maximize the transmission rate at the RUs while maximizing the localization error of MUs. This is achieved by optimizing the beamforming vector, the proportion and reflection precoding of the ARIS, and the AN vector.

According to (\ref{crlb}), the CRLB of the MU localization model is approximately proportional to the square of the ISR at MUs. Therefore, instead of directly maximizing the localization error, we can reformulate the optimization objective to maximize the ISR at the MUs. The specific optimization objective can be expressed as
\begin{subequations}\label{p}
	\begin{align}
		\mathcal{P}_0 :& \max_{\mathbf{w}, \mathbf{\Theta}, \mathbf{v}_e, \boldsymbol{\rho}, \boldsymbol{\eta}}  Q_1 = \sum_{k=1}^{K} \log_2 (1 + \gamma_k) + \omega\sum_{e=1}^{E}\kappa_e, \\
		\text{s.t.} \ &\max_{\mathbf{\Theta}_e}\frac{\left| \mathbf{g}_{e,e}^\dagger \mathbf{\Theta}_e \mathbf{v}_e\right|^2}{\sum_{k=1}^{K}\left|\left(\mathbf{g}_{e,0}^\dagger \mathbf{\Theta} \mathbf{H} + \mathbf{g}_{e,e}^\dagger \mathbf{\Theta}_e \mathbf{H}_e\right) \mathbf{w}_k\right|^2}\geq {\omega}\kappa_{st} ,\forall e,\label{cc1}\\
		&\max_{\mathbf{\Theta}}\frac{\left|\mathbf{g}_{k,0}^\dagger \mathbf{\Theta} \mathbf{H} \mathbf{w}_k\right|^2}{\left\| \mathbf{g}_{k,0}^\dagger \mathbf{\Theta} \right\|^2 \sigma_v^2 + \sigma^2}\geq\gamma_{st},\forall k,\label{c0}\\
		&\varrho_e \geq \varrho_{st},\forall e,\label{cc4}\\
		&\sum_{k=1}^{K} \|\mathbf{w}_k\|^2 \leq P_{S}^{\max},\label{cc2}\\
		&\sum_{k=1}^{K} \|\mathbf{\Theta} \mathbf{H} \mathbf{w}_k\|^2 + \|\mathbf{\Theta}\|_F^2 \sigma_v^2 \leq P_0 = \eta_0 P_R^{\max},\label{cc3}\\
		&\sum_{k=1}^{K}\left\|\mathbf{\Theta}_e \mathbf{H}_e\mathbf{w}_k\right\|^2 + \left\| \mathbf{\Theta}_e \mathbf{v}_e\right\|^2 \leq P_e = \eta_e P_R^{\max},\label{c4}\\
		& \rho_0 + \sum_{e=1}^{E} \rho_e = 1,\label{c5}\\
		&\eta_0 + \sum_{e=1}^{E} \eta_e = 1\label{c6},
	\end{align}
\end{subequations}
where $\gamma_{st}$ represents the minimum transmission rate for each RU, while $\kappa_{st}$ represents the minimum ISR for each MU. Furthermore, we introduce a weight factor $\omega$ to adjust the balance between the two optimization objectives. Additionally, $P_S^{\max}$ and $P_R^{\max}$ represent the maximum power constraints for the SU and the ARIS, respectively. Additionally, constraints (\ref{cc1}) and (\ref{c0}) are the constraints on the signal reflected by ARIS and the noise it generates. Furthermore, According to (\ref{crlb}), we introduce a lower bound constraint in (\ref{cc4}) on the expectation of the ISR at any MU to ensure interference with the localization.
\vspace{-5mm}
\subsection{ARIS Partitioning and Power Allocation Optimization}
In this paper, we primarily focus on solving the optimal partition ratio and power allocation under average channel conditions. To address optimization problem (\ref{p}), we can push the transmission rate of the RUs and the ISR of the MUs to their respective extreme values, thereby decomposing the original optimization problem into two subproblems for separate solutions. Then, we will individually solve the optimal partition and power allocation for ARIS-CE and ARIS-LI.

\textit{1) Optimal Power Allocation for ARIS:} To ensure effective localization interference, the relevant subproblem can be decoupled from (\ref{p}), which is formulated as
\refstepcounter{equation}
\begin{align}
	& \frac{P_{v,e}d_{g,e}^{-\epsilon }}{P_Sd_{h,e}^{-\epsilon } + P_S\left(\sum_{n=1}^{N_0}p_n^2 + N_e\right)L_0\left(d_{g,e}d_{H}\right)^{-\epsilon}} \geq \varrho_{st}\tag{\theequation a}\label{all}\\
	&\text{s.t.}\sum_{k=1}^{K} \|\mathbf{\Theta} \mathbf{H} \mathbf{w}_k\|^2 + \|\mathbf{\Theta}\|_F^2 \sigma_v^2 \leq P_0 = \eta_0 P_R^{\max},\tag{\theequation b}\label{ccc3}\\
	&\sum_{k=1}^{K}\left\|\mathbf{\Theta}_e \mathbf{H}_e\mathbf{w}_k\right\|^2 + \left\| \mathbf{\Theta}_e \mathbf{v}_e\right\|^2 \leq P_e = \eta_e P_R^{\max},\tag{\theequation c}\label{ccc4}\\
	&\eta_0 + \sum_{e=1}^{E} \eta_e = 1\tag{\theequation d}.
\end{align}

Take the upper bound values of constraints (\ref{ccc3}) and (\ref{ccc4}) under the average channel conditions, respectively, which can be reformulated as
\begin{equation}
	\sum_{n=1}^{N_0}p_n^2 = \frac{P_0}{L_{H}P_S + \sigma_{v}^2}, P_{v,e}= P_e - N_eP_SL_{H}.\label{pp}
\end{equation}

Furthermore, combining (\ref{pp}), we can obtain the boundary value of (\ref{all}), which is expressed as
\begin{equation}
	\frac{P_e - N_eP_SL_{H}}{P_S\left(\frac{d_{g,e}}{d_{h,e}}\right)^\epsilon+P_S\left(\frac{P_0}{L_{H}P_S + \sigma_{v}^2}+N_e\right)L_H} = \varrho_{st}.
\end{equation}

Meanwhile, due to the negligible magnitude of $N_eL_H$ and $\sigma_{v}^2$ compared to other dominant terms, their contribution to the overall expression is negligible. Additionally, let $V_e = \frac{d_{g,e}}{d_{h,e}}$. Thus, the boundary value of (\ref{all}) can be further simplified as
\begin{equation}
	\frac{\eta_eP^{\max}_R}{P_SV_e^\epsilon+\eta_0P^{\max}_R} = \varrho_{st}, \eta_0 + \sum_{e=1}^{E} \eta_e = 1.\label{alll}
\end{equation}

By solving (\ref{alll}), we obtain the optimal power allocation for ARIS, which can be expressed as
\begin{equation}\label{pal}
	\begin{split}
		&\eta_0=\frac{P^{\max}_R-\varrho_{st}P_S\sum_{e=1}^{E}V_e^\epsilon}{P_R^{\max}\left(1+\varrho_{st}E\right)},\\ &\eta_e = \varrho_{st}\left(\frac{P_S}{P_{\max}^R}V_e^\epsilon+\eta_0\right).
	\end{split}
\end{equation}

\textit{2) Optimal Partition for ARIS-CE:} To facilitate solving for the optimal solution under average channel conditions, we assume that the interference between RUs can be effectively mitigated or eliminated by designing a proper beamforming vector. Then, a subproblem can be extracted from optimization problem (\ref{p}), which is formulated as
\begin{subequations}
	\begin{align}
		\mathcal{P}_2 : & \min_{k\in \mathcal{K}}\left\{\max_{\mathbf{\Theta}}\frac{\left|\mathbf{g}_{k,0}^\dagger \mathbf{\Theta} \mathbf{H}\mathbf{w}_k\right|^2}{ \left\| \mathbf{g}_{k,0}^\dagger \mathbf{\Theta} \right\|^2 \sigma_v^2 + \sigma^2}\right\}\geq\gamma_{st}, \label{kopt}\\
		\text{s.t.} ~~~	&\sum_{k=1}^{K}\left\|\mathbf{\Theta} \mathbf{H}\mathbf{w}_k\right\|^2 + \sum_{n=1}^{N_0}p_n^2\sigma_{v}^2 \leq P_0\label{c32}.
	\end{align}
\end{subequations}

First, solving the expectation of (\ref{kopt}) under average channel conditions, which can be expressed as
\begin{equation}\label{fenmu1}
	\begin{split}
		\mathbb{E}\left\{\left\| \mathbf{g}_k^\dagger \mathbf{\Theta} \right\|^2 \sigma_v^2 + \sigma^2\right\}= \sum_{n=1}^{N_0}p_n^2L_{g,k}\sigma_{v}^2+\sigma^2.
	\end{split}
\end{equation}

According to (\ref{fenmu1}), the denominator term in (\ref{kopt}) is independent of $\mathbf{\Theta}$ a under the expectation of average channel conditions. Therefore, we  only need to adjust $\mathbf{\Theta}$ for maximizing the numerator term of (\ref{kopt}) under ideal conditions:
\begin{equation}
	\begin{split}
		&\mathbb{E}\left[\max_{\mathbf{\Theta}}\left|\mathbf{g}_{k,0}^\dagger \mathbf{\Theta} \mathbf{H}\mathbf{w}_k\right|^2\right]=\frac{\pi^2}{16}N_0\sum_{n=1}^{N_0}p_n^2P_kL_{g,k}L_{H} ,\label{fenzi1}
	\end{split}
\end{equation}

%\begin{equation}\label{fenmu1}
%	\begin{split}
%		\mathbb{E}\left\{\left\| \mathbf{g}_k^\dagger \mathbf{\Theta} \right\|^2 \sigma_v^2 + \sigma^2\right\}= N_0p^2L_{g,k}\sigma_{v}^2+\sigma^2
%	\end{split}
%\end{equation}
%\begin{equation}
%	\mathbb{E}\left\{\sum_{a=1, a \neq k}^{K} |\overline{\mathbf{h}}_k^\dagger \mathbf{w}_a|^2 + \left\| \mathbf{g}_k^\dagger \mathbf{\Theta} \right\|^2 \sigma_v^2 + \sigma^2\right\} = N_0p^2\chi_{g,k}^2\sigma_{v}^2+\sigma^2, \label{fenmu1}
%\end{equation}

Substituting into (\ref{kopt}), then the optimization problem for partition of ARIS-CE can be reformulated  as 
\begin{equation}
	\frac{\pi^2}{16}\frac{N_0 P_k P_0L_{g,k}L_{H}}{ P_0L_{g,k}\sigma_v^2+P_SL_{H}\sigma^2+\sigma^2\sigma_v^2}\geq\gamma_{st}.\label{gammamin}
\end{equation}

Then, pushing $\gamma_k$ to its minimum subject in (\ref{gammamin}), we obtain the optimal partition for ARIS-CE, which is expressed as

%\begin{strip}
%	\rule{\textwidth}{.4pt}
%	\begin{equation}
%		\gamma_k=\frac{16P_kL_{h,k}\left(L_{H}P_S+\sigma_v^2\right)+\pi^2 N_0 P_k P_0L_{g,k}L_{H}}{\left(P_S - P_k\right)\left[16L_{h,k}\left(L_{H}P_S+\sigma_v^2\right)+\pi^2 N_0 P_0L_{g,k}L_{H}\right] + 16\left(P_0L_{g,k}\sigma_v^2+P_SL_{H}\sigma^2+\sigma^2\sigma_v^2\right)}\geq\gamma_{st},\label{gammamin}
%	\end{equation}
%\end{strip} 

\begin{equation}
	N_0 = \left\lceil\max_{k\in\mathcal{K}}\left\{\frac{16\gamma_{st}\left(P_0L_{g,k}\sigma_v^2+P_SL_{H}\sigma^2+\sigma^2\sigma_v^2\right)}{\pi^2L_{g,k}L_{H}P_0P_S}\right\}\right\rceil.
\end{equation}

\textit{2) Optimal Partition for ARIS-LI:} Similarly, under average channel conditions, a subproblem can be extracted from (\ref{p}) for ARIS-LI, which is formulated as
\begin{subequations}
	\begin{align}
		\mathcal{P}_3 : &\frac{\left| \mathbf{g}_{e,e}^\dagger \mathbf{\Theta}_e \mathbf{v}_e\right|^2}{\sum_{k=1}^{K}\left|\left(\mathbf{g}_{e,0}^\dagger \mathbf{\Theta} \mathbf{H} + \mathbf{g}_{e,e}^\dagger \mathbf{\Theta}_e \mathbf{H}_e\right) \mathbf{w}_k\right|^2}\geq\kappa_{st}^*, \label{gopt}\\
		\text{s.t.} \ &\sum_{k=1}^{K}\left\|\mathbf{\Theta}_e \mathbf{H}_e\mathbf{w}_k\right\|^2 + \left\| \mathbf{\Theta}_e \mathbf{v}_e\right\|^2 \leq P_e\label{cc21}.
	\end{align}
\end{subequations}

First, solving the expectation of (\ref{gopt}) under average channel conditions, which can be expressed as
\begin{equation}
	\begin{split}
		\mathbb{E}&\left[\max_{\boldsymbol{\theta}_e}\left| \mathbf{g}_{e,e}^\dagger \mathbf{\Theta}_e \mathbf{v}_e\right|^2\right]=\frac{\pi }{4}N_eL_{g,e}P_{v,e}.\label{fenzi2}
	\end{split}
\end{equation}
\begin{equation}\label{fenmu2}
	\begin{split}
		&\mathbb{E}\left[\sum_{k=1}^{K}\left|\left(\mathbf{g}_{e,0}^\dagger \mathbf{\Theta} \mathbf{H} + \mathbf{g}_{e,e}^\dagger \mathbf{\Theta}_e \mathbf{H}_e\right) \mathbf{w}_k\right|^2\right]\\
		&=P_S L_{g,e}L_{H}\left(1+\frac{P_0}{L_{H}P_S + \sigma_{v}^2}\right).
	\end{split}
\end{equation}

Substitute (\ref{pp}) into (\ref{fenmu2}) and (\ref{fenzi2}), then we can obtain
\begin{equation}
	-P_SL_HN_e^2+P_eN_e-C\geq 0.
\end{equation}
\begin{equation}
	C = \frac{4}{\pi}\kappa_{st}^*L_H\left(1+\frac{P_0}{L_{H}P_S + \sigma_{v}^2}\right).
\end{equation}

Solving the above inequality, we can obtain the number of elements in the ARIS-LI partition under ideal conditions, which is formulated as
\begin{equation}
	N_e=\left\lceil\frac{P_e - \sqrt{P_e^2 - 4P_SL_H C}}{2P_SL_H}\right\rceil.
\end{equation}

Furthermore, the ARIS needs to perform localization interference on MUs with the premise of ensuring enhanced communication for RUs. For ease of representation, let $N_E = \sum_{e=1}^{E}N_e$. Therefore, the final ARIS partition decision must satisfy the following principles:
\begin{equation}\label{final}
	\begin{split}
		&N_0^*=\left\{\begin{matrix}
				&N_0&\left(N_0+N_E \geq N_t\right),\\
				&N_t-N_E&\mbox{else},
			\end{matrix}\right.\\
	    &N_e^*=\left\{\begin{matrix}
	    		&\frac{N_e}{N_E}\left(N_t-N_0\right),&\left(N_0+N_E \geq N_t\right),\\
	    		&N_e&\mbox{else}.
	    	\end{matrix}\right.
	\end{split}
\end{equation}

\subsection{Joint Optimization Scheme for Beamforming, Reflection Precoding of the ARIS-CE}
Through ARIS partitioning, we can physically decouple the two optimization objectives, communication enhancement and localization interference, allowing us to optimize them separately. For ARIS-CE, the decoupled subproblem from (\ref{p}) can be formulated as
\begin{subequations}\label{p1}
	\begin{align}
		\mathcal{P}_4 : \max_{\mathbf{w}, \mathbf{\Theta}} &Q_2 = \sum_{k=1}^{K} \log_2 (1 + \gamma_k), \\
		\text{s.t.} \ &\sum_{k=1}^{K} \|\mathbf{w}_k\|^2 \leq P_{S}^{\max},\\
		&\sum_{k=1}^{K} \|\mathbf{\Theta} \mathbf{H} \mathbf{w}_k\|^2 + \|\mathbf{\Theta}\|_F^2 \sigma_v^2 \leq P_0.
	\end{align}
\end{subequations}

The non-convexity of (\ref{p1}) makes it challenging to solve directly. To handle this non-convex logarithmic and fractional problem, we employ the FP method to decouple (\ref{p1}), enabling the optimization of multiple variables separately. Therefore, the following lemma needs to be introduced \cite{shen2018fractional}:

\begin{lemma}\label{l1}
	By introducing auxiliary variables $\boldsymbol{\varepsilon}\triangleq \left[\varepsilon_1,\cdots,\varepsilon_K\right]$ and $\boldsymbol{\Upsilon}\triangleq \left[\Upsilon_1,\cdots,\Upsilon_K\right]$ into (\ref{p1}), it can be equivalently transformed into
	\begin{subequations}\label{p4}
		\begin{align}
			\mathcal{P}_5 : &\max_{\mathbf{w}, \mathbf{\Theta}, \boldsymbol{\varepsilon}, \boldsymbol{\Upsilon}} Q_3 = \sum_{k=1}^{K} \left[\log_2 \left(1 + \varepsilon_k\right)-\varepsilon_k\right]\notag\\	
			+\sum_{k=1}^{K}&\left[2\sqrt{1+\varepsilon_k} \mathcal{R}\left\{\Upsilon^*\overline{\mathbf{h}}_k^\dagger \mathbf{w}_k\right\} \right]\notag\\
			-\sum_{k=1}^{K}& |\Upsilon_k|^2\left(\sum_{a=1}^{K} |\overline{\mathbf{h}}_k^\dagger \mathbf{w}_a|^2 + \left\| \mathbf{g}_{k,0}^\dagger \mathbf{\Theta} \right\|^2 \sigma_v^2 + \sigma^2\right),\\
			\text{s.t.} \ &\sum_{k=1}^{K} \|\mathbf{w}_k\|^2 \leq P_{S}^{\max},\label{ccc1}\\
			&\sum_{k=1}^{K} \|\mathbf{\Theta} \mathbf{G} \mathbf{w}_k\|^2 + \|\mathbf{\Theta}\|_F^2\sigma_v^2 \leq P_0.\label{c2}
		\end{align}
	\end{subequations}
\end{lemma}

$\mathit{Proof.}$ Detailed proof can be found in \cite[Subsection III]{shen2018fractional}.

$\textbf{Remarks}$: Lemma \ref{l1} decouples the original joint optimization problem into an alternate optimization of the SU beamforming vector $\mathbf{w}$, ARIS-CE reflection precoding $\boldsymbol{\Theta}$, auxiliary variables $\boldsymbol{\varepsilon}$ and $\boldsymbol{\Upsilon}$. According to \cite{shen2018fractional}, if all variables in (\ref{p4}) are optimal in each iteration update, a locally optimal solution to (\ref{p4}) can be obtained after convergence. 

According to Lemma \ref{l1}, we will provide a detailed introduction to the optimization steps for each variable, and summarize the proposed joint optimization scheme in Algorithm \ref{al3}.
\begin{algorithm}[!t]
	\caption{Joint Optimization Scheme for Communication Enhancement and Localization Interference of ARIS.}
	\label{al3}
	\begin{algorithmic}[1]
		\STATE \textbf{Input:} Channel matrices $\mathbf{h}_k\in \mathbb{C}^{M\times 1}$, $\mathbf{h}_e\in \mathbb{C}^{M\times 1}$, $\mathbf{H}\in \mathbb{C}^{N_0\times M}$, $\mathbf{H}_e\in \mathbb{C}^{N_e\times M}$ $\mathbf{g}_{k, 0}\in \mathbb{C}^{N_0\times 1}$, $\mathbf{g}_{k, e}\in \mathbb{C}^{N_e\times 1}$, $\mathbf{g}_{e, 0}\in \mathbb{C}^{N_0\times 1}$ and $\mathbf{g}_{e, i}\in \mathbb{C}^{N_i\times 1}$ for all $k \in \mathcal{K}$ and all $e \in \mathcal{E}$. Position $\mathbf{l}_S\triangleq\left[x_S\ y_S \ z_S\right]^\top$, $\mathbf{l}_{R}\triangleq\left[x_{R}\ y_{R} \ z_{R}\right]^\top$, $\left\{\mathbf{l}_{r,k} \triangleq\left[x_{r,k}\ y_{r,k} \ z_{r,k}\right]^\top, k\in\mathcal{K}\right\}$, and $\left\{\mathbf{l}_e\triangleq\left[x_e\ y_e \ z_e\right]^\top, e \in\mathcal{E}\right\}$.
		\STATE \textbf{Output:} SU beamforming vector $\mathbf{w}$, partition $\boldsymbol{\rho}$ and power allocation $\boldsymbol{\eta}$ of ARIS, reflection precoding of ARIS-CE $\boldsymbol{\Theta}$, reflection precoding of ARIS-LI $\boldsymbol{\Theta}_e$, and AN factor $\mathbf{v}_e$,
		\STATE Calculate power allocation $\boldsymbol{\eta}$ of ARIS by (\ref{alll});
		\STATE Calculate partition $\boldsymbol{\rho}$ of ARIS by (\ref{final});
		\STATE Initialize $\mathbf{w}$, $\boldsymbol{\Theta}$, $\boldsymbol{\Theta}_e$ and $\mathbf{v}_e$.
		\WHILE{no convergence of $Q_3$}
		\STATE Fix $\left(\mathbf{w},\boldsymbol{\Theta},\boldsymbol{\Upsilon}\right)$, and update $\boldsymbol{\varepsilon}$ by (\ref{zeta});
		\STATE Fix $\left(\mathbf{w},\boldsymbol{\Theta},\boldsymbol{\varepsilon}\right)$, and update $\boldsymbol{\Upsilon}$ by (\ref{chi});
		\STATE Fix variables $\left(\boldsymbol{\Theta},\boldsymbol{\varepsilon},\boldsymbol{\Upsilon}\right)$, and update the SU beamforming vector $\mathbf{w}$ by (\ref{w1});
		\STATE Fix $\left(\mathbf{w},\boldsymbol{\varepsilon},\boldsymbol{\Upsilon}\right)$, and update the reflection precoding of ARIS-CE $\boldsymbol{\Theta}$ by (\ref{theta1})
		\ENDWHILE
		\WHILE{no convergence of $Q_4$}
		\STATE Fix $\left(\mathbf{v}_e,\boldsymbol{\Theta}_e\right)$, and update $\xi$ by (\ref{xi});
		\STATE Fix $\left(\boldsymbol{\Theta}_e,\xi\right)$, and update AN vector $\mathbf{v}_e$ by (\ref{ve});
		\STATE Fix $\left(\mathbf{v}_e,\xi\right)$, and update the reflection precoding of ARIS-LI $\boldsymbol{\Theta}_e$ by solving SDP problem (\ref{p11});
		\ENDWHILE
	\end{algorithmic}
\end{algorithm}

\text{1)} Optimize $\boldsymbol{\varepsilon}$: Fix variables $\left(\mathbf{w},\boldsymbol{\Theta},\boldsymbol{\Upsilon}\right)$, and optimize auxiliary variables $\boldsymbol{\varepsilon}$ as
\begin{equation}\label{zeta}
	\varepsilon_k^{op} = \frac{|\overline{\mathbf{h}}_k^\dagger \mathbf{w}_k|^2\sigma_v^2}{\sum_{a=1}^{K} |\overline{\mathbf{h}}_k^\dagger \mathbf{w}_a|^2 + \left\| \mathbf{g}_{k,0}^\dagger \mathbf{\Theta} \right\|^2 \sigma_v^2 + \sigma^2}.
\end{equation} 

\text{2)} Optimize $\boldsymbol{\Upsilon}$: Fix variables $\left(\mathbf{w},\boldsymbol{\Theta},\boldsymbol{\varepsilon}\right)$, and optimize auxiliary variables $\boldsymbol{\Upsilon}$ by solving $\partial Q_2 / \partial \Upsilon_k=0$ as
\begin{equation}\label{chi}
	\Upsilon_k^{op} = \frac{\sqrt{1+\varepsilon_k}\overline{\mathbf{h}}_k^\dagger \mathbf{w}_k}{\sum_{a=1}^{K} |\overline{\mathbf{h}}_k^\dagger \mathbf{w}_a|^2 + \left\| \mathbf{g}_{k,0}^\dagger k \mathbf{\Theta} \right\|^2\sigma_v^2 + \sigma^2}.
\end{equation}

\text{3)} Optimize $\mathbf{w}$: For shorthand, we further define
\begin{subequations}\label{w}
	\begin{align}
		&\boldsymbol{\alpha}_k \!\!=\!\! \sqrt{1\!+\!\varepsilon_k}\Upsilon_k\overline{\mathbf{h}}_k^\dagger, \boldsymbol{\alpha}\!\!\triangleq\!\!\left[\boldsymbol{\alpha}_1^\top\!,\!\cdots\!,\!\boldsymbol{\alpha}_1^\top\right]^\top\!\!\!,\\
		&\mathbf{B}\!\!=\!\!\mathbf{I}_K\otimes\sum_{k=1}^K |\Upsilon_k|^2 \overline{\mathbf{h}}_k\overline{\mathbf{h}}_k^\dagger, \mathbf{C}\!\!=\!\!\mathbf{I}_K\otimes(\mathbf{H}^\dagger\boldsymbol{\Theta}^\dagger\boldsymbol{\Theta}\mathbf{H}),\\
		&P_w = P_0-\|\mathbf{\Theta}\|_F^2 \sigma_v^2.
	\end{align}
\end{subequations}

Then, fix variables $\left(\boldsymbol{\Theta},\boldsymbol{\varepsilon},\boldsymbol{\Upsilon}\right)$, and obtain a new optimization problem for the SU beamforming vector $\mathbf{w}$ based on (\ref{p4}), which can be formulated as
\begin{subequations}\label{p5}
	\begin{align}
		\mathcal{P}_6 : \max_{\mathbf{w}}\  &\mathcal{R}\left\{2\boldsymbol{\alpha}^\dagger\mathbf{w}\right\}-\mathbf{w}^\dagger\mathbf{B}\mathbf{w}, \\
		\text{s.t.} &\left\| \mathbf{w}\right\|^2 \leq P_{S}^{\max},\\
		&\mathbf{w}^\dagger\mathbf{C}\mathbf{w} \leq P_w.
	\end{align}
\end{subequations} 

As a standard quadratic constraint quadratic programming, optimization problem (\ref{p5}) can be solved by applying the Lagrange multiplier method. Specifically, we construct the Lagrange function for (\ref{p5}), which is expressed as 
\begin{equation}
	\begin{split}
		&\mathcal{L}(\mathbf{w},\lambda_1, \lambda_2)=\mathcal{R}\left\{2\boldsymbol{\alpha}^\dagger\mathbf{w}\right\}-\mathbf{w}^\dagger\mathbf{B}\mathbf{w}\\
		&+\lambda_1\left(P_{S}^{\max}-\mathbf{w}^\dagger\mathbf{w}\right) + \lambda_2\left(P_w - \mathbf{w}^\dagger\mathbf{C}\mathbf{w}\right),
	\end{split}
\end{equation}
where $\lambda_1$ and $\lambda_2$ are the Lagrange multipliers. Then, taking the derivative of the Lagrange function $\mathcal{L}(\mathbf{w},\lambda_1, \lambda_2)$ with respect to $\mathbf{w}$ and set the derivative to zero, the solution for $\mathbf{w}$ can be obtained as
\begin{equation}\label{w1}
	\mathbf{w}_{op}=\left(\mathbf{{B} + \lambda_1\mathbf{I}_{KM}+\lambda_2\mathbf{C}}\right)^{-1}\boldsymbol{\alpha},
\end{equation}
wherein the optimal Lagrange multiplier $\lambda_1$ and $\lambda_2$ that satisfy constraints can be obtained through a binary search \cite{boyd2011distributed}. 

\text{4)} Optimize $\boldsymbol{\Theta}$: For shorthand, we further define
\begin{subequations}
	\begin{align}
		\boldsymbol{\upsilon} = &\sum_{k=1}^{K} \sqrt{1\!+\!\varepsilon_k} \mathrm{Diag} \left( \boldsymbol{\Upsilon}_k^* \mathbf{g}_{k,0}^\dagger \right) \mathbf{H} \mathbf{w}_k  \notag\\
		&-\sum_{k=1}^{K} |\boldsymbol{\Upsilon}_k|^2 \mathrm{Diag} (\mathbf{g}_{k,0}^\dagger) \mathbf{H} \sum_{a=1}^{K} \mathbf{w}_a \mathbf{w}_a^\dagger \mathbf{h}_k,\\
		\boldsymbol{\Lambda} = &\sum_{k=1}^{K} |\boldsymbol{\Upsilon}_k|^2 \sum_{j=1}^{K} \mathrm{Diag} \left( \mathbf{g}_{k,0}^\dagger \right) \mathbf{H} \mathbf{w}_j \mathbf{w}_j^\dagger \mathbf{H}^\dagger \mathrm{Diag} \left( \mathbf{g}_{k,0} \right) \notag\\
		&+\sum_{k=1}^{K} |\boldsymbol{\Upsilon}_k|^2 \mathrm{Diag} \left( \mathbf{g}_{k,0}^\dagger \right) \mathrm{Diag} \left( \mathbf{g}_{k,0} \right) \sigma_v^2,\\
		\boldsymbol{\Psi} = &\sum_{k=1}^{K} \mathrm{Diag}(\mathbf{H} \mathbf{w}_k) \left( \mathrm{Diag}(\mathbf{H} \mathbf{w}_k) \right)^\dagger + \sigma_v^2 \mathbf{I}_M.
	\end{align}
\end{subequations}

Then, fix variables $\left(\mathbf{w},\boldsymbol{\varepsilon},\boldsymbol{\Upsilon}\right)$, and obtain a new optimization problem for the reflection precoding $\boldsymbol{\Theta}$ of ARIS-CE based on (\ref{p4}), which can be formulated as
\begin{subequations}\label{p6}
	\begin{align}
		\mathcal{P}_7 : \max_{\boldsymbol{\theta}}\  &\mathcal{R}\left\{2\boldsymbol{\theta}^\dagger\boldsymbol{\upsilon}\right\}-\boldsymbol{\theta}^\dagger\boldsymbol{\Lambda}\boldsymbol{\theta}, \\
		\text{s.t.} \ & \boldsymbol{\theta}^\dagger\boldsymbol{\Psi}\boldsymbol{\theta} \leq P_0.\label{con}
	\end{align}
\end{subequations}

Furthermore, we use the Lagrange multiplier method to solve optimization problem (\ref{p6}). Specifically, we introduce the Lagrange multiplier $\mu$ and construct the Lagrange function as 
\begin{equation}
	\mathcal{L}(\boldsymbol{\theta},\mu)=\mathcal{R}\left\{2\boldsymbol{\theta}^\dagger\boldsymbol{\upsilon}\right\}-\boldsymbol{\theta}^\dagger\boldsymbol{\Lambda}\boldsymbol{\theta}+\mu(P_0-\boldsymbol{\theta}^\dagger\boldsymbol{\Psi}\boldsymbol{\theta}).
\end{equation}

Then, taking the derivative of the Lagrange function $\mathcal{L}(\boldsymbol{\theta},\mu)$ with respect to $\boldsymbol{\theta}$ and set the derivative to zero, the solution for $\boldsymbol{\theta}$ can be obtained as
\begin{equation}\label{theta1}
	\boldsymbol{\theta}=\left(\boldsymbol{\Lambda}+\mu \boldsymbol{\Psi}\right)^{-1}\boldsymbol{\upsilon},
\end{equation}
wherein the optimal Lagrange multiplier $\mu$ that satisfies power constraint (\ref{con}) can be obtained through a binary search \cite{boyd2011distributed}.
\subsection{Joint Optimization Scheme for Reflection Precoding and Artificial Noise of the ARIS-LI}
For ARIS-LI, the decoupled subproblem from (\ref{p}) can be formulated as
\begin{subequations}\label{p7}
	\begin{align}
		\mathcal{P}_8 : \max_{\mathbf{\Theta}_e, \mathbf{v}_e} &\kappa_e= \frac{\left| \mathbf{g}_{e,e}^\dagger \mathbf{\Theta}_e \mathbf{v}_e\right|^2}{\sum_{k=1}^{K}\left|\left(\overline{\mathbf{h}}_e^\dagger + \mathbf{g}_{e,e}^\dagger \mathbf{\Theta}_e \mathbf{H}_e \right)\mathbf{w}_k\right|^2},\\
		\text{s.t.} \ &\sum_{k=1}^{K}\left\| \mathbf{H}_e\mathbf{w}_k\right\|^2 + \left\| \mathbf{v}_e\right\|^2 \leq P_e.
	\end{align}
\end{subequations}

Similar to solving (\ref{p1}), we utilize the FP method to decouple (\ref{p7}), allowing for separation of multiple variable optimization. Likewise, the following lemma is introduced to facilitate the subsequent solution:

\begin{lemma}\label{l2}
	By introducing auxiliary variables $\xi$ and into (\ref{p7}), it can be equivalently transformed into
	\begin{subequations}\label{p8}
		\begin{align}
			\mathcal{P}_9 : \max_{\mathbf{w}, \mathbf{\Theta}, \xi} Q_4 &= 	
			2 \mathcal{R}\left\{\xi^*\mathbf{g}_{e,e}^\dagger \mathbf{\Theta}_e \mathbf{v}_e\right\} \notag\\
			-& |\xi|^2\sum_{k=1}^{K}\left|\left(\overline{\mathbf{h}}_e^\dagger + \mathbf{g}_{e,e}^\dagger \mathbf{\Theta}_e \mathbf{H}_e \right)\mathbf{w}_k\right|^2,\\
			\text{s.t.} \ &\sum_{k=1}^{K}\left\| \mathbf{H}_e\mathbf{w}_k\right\|^2 + \left\| \mathbf{v}_e\right\|^2 \leq P_e.
		\end{align}
	\end{subequations}
\end{lemma}

$\mathit{Proof.}$ Detailed proof can be found in \cite[Subsection II]{shen2018fractional}.

$\textbf{Remarks}$: Lemma \ref{l2} decouples the original joint optimization problem into an alternate optimization of ARIS-LI reflection precoding $\boldsymbol{\Theta}_e$, AN $\mathbf{v}_e$, and auxiliary variables $\xi$.  

According to Lemma \ref{l2}, the optimization steps for each variable are summarized as follows.

\text{1)} Optimize $\xi$: Fix variables $\left(\mathbf{v}_e,\boldsymbol{\Theta}_e\right)$, and optimize auxiliary variables $\xi$ by solving $\partial Q_3 / \partial \xi=0$ as
\begin{equation}\label{xi}
	\xi^{op} = \frac{\left| \mathbf{g}_{e,e}^\dagger \mathbf{\Theta}_e \mathbf{v}_e\right|^2}{\sum_{k=1}^{K}\left|\left(\overline{\mathbf{h}}_e^\dagger + \mathbf{g}_{e,e}^\dagger \mathbf{\Theta}_e \mathbf{H}_e \right)\mathbf{w}_k\right|^2}.
\end{equation}

\text{2)} Optimize $\mathbf{v}_e$: It can be observed that in $Q_3$, only the numerator term is related to $\mathbf{v}_e$. Therefore, for $\mathbf{v}_e$, maximizing $Q_3$ is equivalent to maximizing its numerator term. For shorthand, we further define
\begin{equation}
	\mathbf{B}_e\!\!=\mathbf{\Theta}_e^\dagger\mathbf{g}_{e,e}\mathbf{g}_{e,e}^\dagger \mathbf{\Theta}_e.
\end{equation}

Then, fix variables $\left(\boldsymbol{\Theta}_e,\xi\right)$, and obtain a new optimization problem for the AN vector $\mathbf{v}_e$ based on (\ref{p8}), which can be formulated as
\begin{subequations}\label{p9}
	\begin{align}
		\mathcal{P}_{10} : \max_{\mathbf{v}_e}\  &\mathbf{v}_e^\dagger\mathbf{B}_e\mathbf{v}_e, \\
		\text{s.t.} \ & \left\|\mathbf{v}_e\right\|^2 \leq P_e-\sum_{k=1}^{K}\left\| \mathbf{H}_e\mathbf{w}_k\right\|^2.
	\end{align}
\end{subequations} 

The above optimization problem is a Rayleigh quotient problem, thus the optimal solution $\mathbf{v}_e^{op}$ is expressed as
\begin{equation}\label{ve}
	\mathbf{v}_e^{op} = \sqrt{P_e-\sum_{k=1}^{K}\left\| \mathbf{H}_e\mathbf{w}_k\right\|^2}\mathbf{u}_{\max},
\end{equation}
where $\mathbf{u}_{\max}$ is the eigenvector corresponding to the largest eigenvalue of $\mathbf{B}_e$.

\text{3)} Optimize $\boldsymbol{\Theta}_e$: Let $\mathbf{G}_e = \mathrm{Diag} \left( \mathbf{g}_{e,e}\right) $. For shorthand, we further define
\begin{subequations}
	\begin{align}
		\boldsymbol{\upsilon}_e = &\xi^*\mathbf{G}_e^\dagger\mathbf{v}_e  \!-\!|\xi|^2 \mathbf{G}_e^\dagger \mathbf{H}_e \sum_{a=1}^{K} \mathbf{w}_a \mathbf{w}_a^\dagger \overline{\mathbf{h}}_e,\\
		\boldsymbol{\Lambda}_e = &|\xi|^2 \mathbf{G}_e^\dagger\mathbf{H}_e \sum_{k=1}^{K}\mathbf{w}_k \mathbf{w}_k^\dagger \mathbf{H}_e^\dagger \mathbf{G}_e.
	\end{align}
\end{subequations}

Then, fix variables $\left(\mathbf{v}_e,\xi\right)$, and obtain a new optimization problem for the ARIS-LI reflection precoding $\boldsymbol{\Theta}_e$, which can be formulated as
\begin{subequations}\label{p10}
	\begin{align}
		\mathcal{P}_{11} : &\max_{\boldsymbol{\theta}_e}\  2\mathcal{R}\left\{\boldsymbol{\theta}_e^\dagger\boldsymbol{\upsilon}_e\right\}-\boldsymbol{\theta}_e^\dagger\boldsymbol{\Lambda}_e\boldsymbol{\theta}_e, \\
		\text{s.t.} \ & \boldsymbol{\theta}_e\in\left\{\left[e^{j\phi_1}, \cdots,e^{j\phi_{N_e}}\right]^\top|\phi_n\in\left(0, 2\pi \right],\forall n\right\}.
	\end{align}
\end{subequations}

Furthermore, we can reorganize  the above optimization problem. Specifically, we construct an auxiliary matrix $\mathbf{D}_e$ and an auxiliary variable $\mathbf{z}_e$, which can be represented as
\begin{equation}
	\mathbf{D}_e=\left[
	\begin{matrix}
		\boldsymbol{\Lambda}_e & \boldsymbol{\upsilon}_e\\
		\boldsymbol{\upsilon}_e^\dagger & 0
	\end{matrix}\right], \mathbf{z}_e = \left[\boldsymbol{\theta}_e^\top, t\right]^\top.
\end{equation}

Then, we can transform optimization problem (\ref{p10}) into a SDP problem, which can be represented as
\begin{subequations}\label{p11}
	\begin{align}
		\mathcal{P}_{12} : &\max_{\boldsymbol{z}_e}\  \boldsymbol{z}_e^\dagger\mathbf{D}_e\boldsymbol{z}_e, \\
		\text{s.t.} \ & \left|{z}_{e,n}\right| = 1,\forall n\in \mathcal{N}_e.
	\end{align}
\end{subequations}

The SDP problem in (\ref{p11}) can ultimately be solved for the optimal reflection precoding $\boldsymbol{\theta}_e^{op}$ using either the interior-point method or the penalty function method.

\section{Simulation Results and Analysis}\label{sim}
\begin{table}[t]
	\caption{Simulation Parameter Setting.}\label{t1}
	\centering
	\tabcolsep=0.02cm
	\renewcommand\arraystretch{1.3}
	\begin{tabular}{l|c}
		\hline
		\hline
		\textbf{~~~~~~~~~~~~~~~~Parameters} & \textbf{Value} \\
		\hline
		%		Number of clients $C$ & 100 \\
		%		Training epochs $T$ & 400 \\
		Number of antennas $M$ & 4  \\
		Number of RUs $K$& 4 \\
		Background noise power $\sigma^2$ & -100dBm \\
		Intrinsic noise power $\sigma_v^2$ & -100dBm \\
		Measurement error \cite{martin2012modeling} $\sigma_{dB}$ & 10dB \\
		Path loss at the reference distance $L_0$ \cite{access2010further} & -37.3dB \\
		Path loss exponent $\epsilon$\cite{access2010further}~~~ & ~~~2.2~~~ \\
		Position of SU $\mathbf{l}_S$ & [50, 50, 50] \\
		Position of ARIS $\mathbf{l}_R$ & [270, 120, 20] \\
		Central position of the RUs $\mathbf{l}_{R}^0$ & [350, 110, 50] \\
		Distance between RU and central position of the RUs $d_{sr}$ & 5m\\
		%		Seabed reflection coefficient $\kappa_b$ \cite{weiss2022semi} & 0.85 \\
		%		Sampling time interval of the MUs $T_s$ & 0.001s \\
		%		Number of sampling points $N$ & 64 \\
		%		Speed of underwater sound $c$ & 1500m/s \\
		%		Seabed depth $h$ & 100m\\
		%		Position of SU $\mathbf{l}_S$ & (200.7m,140.6m,50.2m)\\
		%		Position of UARIS $\mathbf{l}_U$ & (500m,210m,30m)\\
		%		Maximum transmit power of the SU $P_S^{max}$ & 30dBm\\
		%		Maximum reflected power of UARIS $P_S^{max}$ & 20dBm \\
		\hline
		\hline
	\end{tabular}
	\label{table1}\vspace{-3mm}
	%	\vspace{-0.5cm}
\end{table}
In this section, we validate the effectiveness of the proposed scheme through five sets of simulation experiments. In the simulation experiments, we set the presence of four legitimate RUs, all randomly distributed on a spherical surface with point $\mathbf{l}_{R}^0 \triangleq\left[x_{R}^0\ y_{R}^0 \ z_{R}^0\right]^\top$ as the center of the sphere and radius $d_{sr}$. Meanwhile, we set the MUs randomly distributed on the hemispherical surface with center at $\mathbf{l}_S$ and radius $d_{se}$. The other simulation parameters are summarized in Table \ref{table1}. To demonstrate the effectiveness of the proposed scheme, we conduct simulations for the following four schemes:
\begin{enumerate}
	\item \textbf{ARIS with Adaptive Virtual Partition:} For an ARIS with adaptive virtual partition, the proposed Algorithm \ref{al3} is used to jointly optimize communication enhancement an localization interference.
	\item \textbf{ARIS with Fixed Virtual Partition:} For an ARIS with fixed virtual partition, the proposed Algorithm \ref{al3} is used to optimize ARIS-CE and ARIS-LI with $\rho_0=\eta_0=0.5$ and $\rho_e=\eta_e = 0.5/E$.
	\item \textbf{ARIS without Virtual Partition:} For an ARIS without virtual partition or AN, the algorithm proposed in \cite{zhang2022active} is used to jointly optimize the beamforming of the SU and the reflection precoding of the ARIS.
	\item \textbf{No ARIS:} Without ARIS, the WMMSE algorithm proposed in \cite{shi2011iteratively} is used to optimize the beamforming of the SU.
\end{enumerate}

%\begin{figure*}[htbp]
%	\centering
%	\begin{minipage}{0.49\linewidth}
%		\centering
%		\includegraphics[width=0.9\linewidth]{L_RATE.png}
%		\caption{Sum-rate of different schemes}
%		\label{fig3a}%文中引用该图片代号
%	\end{minipage}
%	%\qquad
%	\begin{minipage}{0.49\linewidth}
%		\centering
%		\includegraphics[width=0.9\linewidth]{L_RMSE.png}
%		\caption{RMSE of different schemes}
%		\label{fig3b}%文中引用该图片代号
%	\end{minipage}
%\end{figure*}

\vspace{-4mm}
\subsection{Coverage Performance of Different Schemes}\label{cov}
\begin{figure}[!t]
	\centering
	\subfigure[Sum-rate of different schemes]{
		\includegraphics[width=0.4\textwidth,height=0.3\textwidth]
		{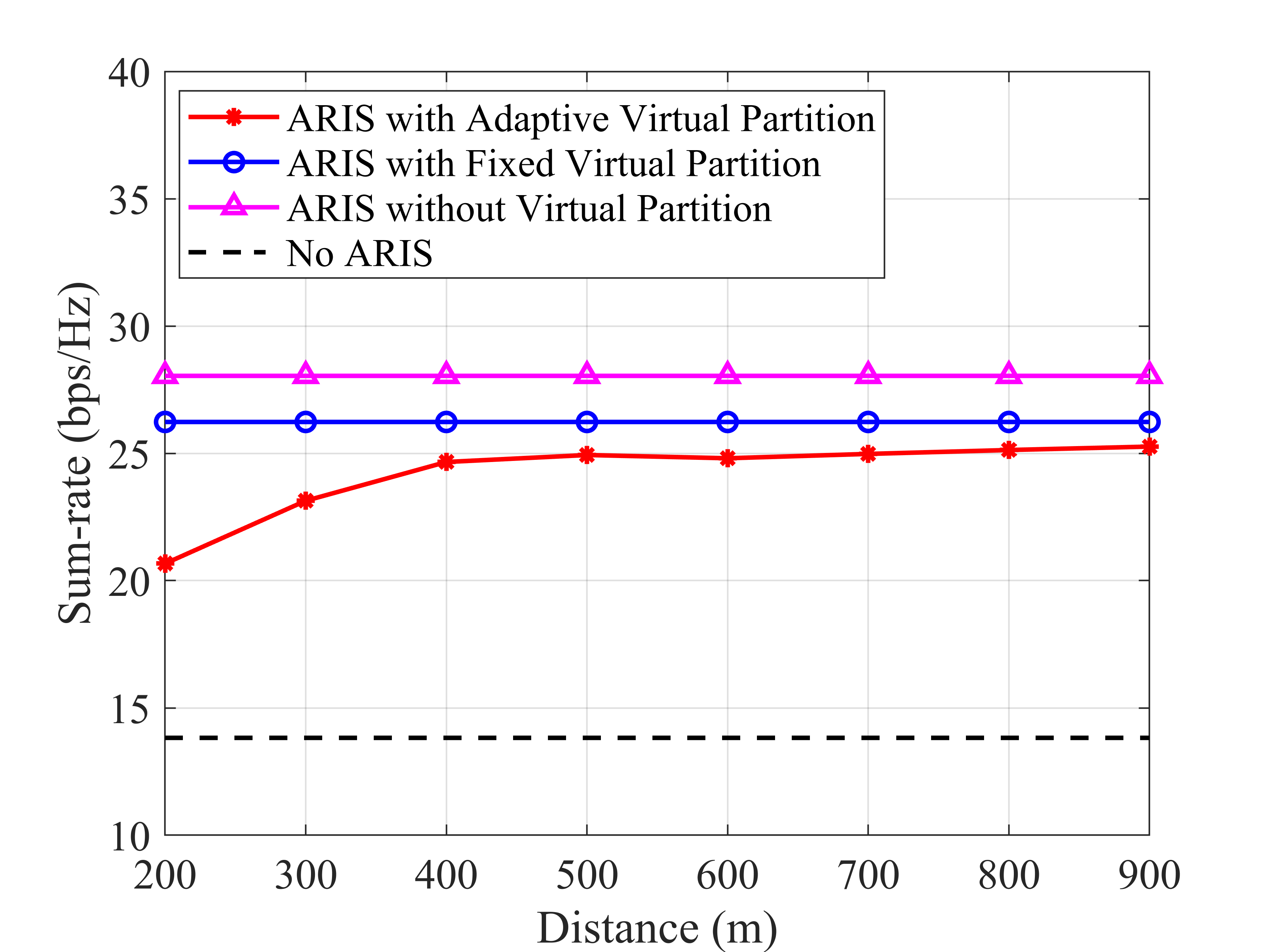}\label{fig3a}
	}
	\centering
	\subfigure[RMSE of different schemes]{
		\includegraphics[width=0.4\textwidth,height=0.3\textwidth]{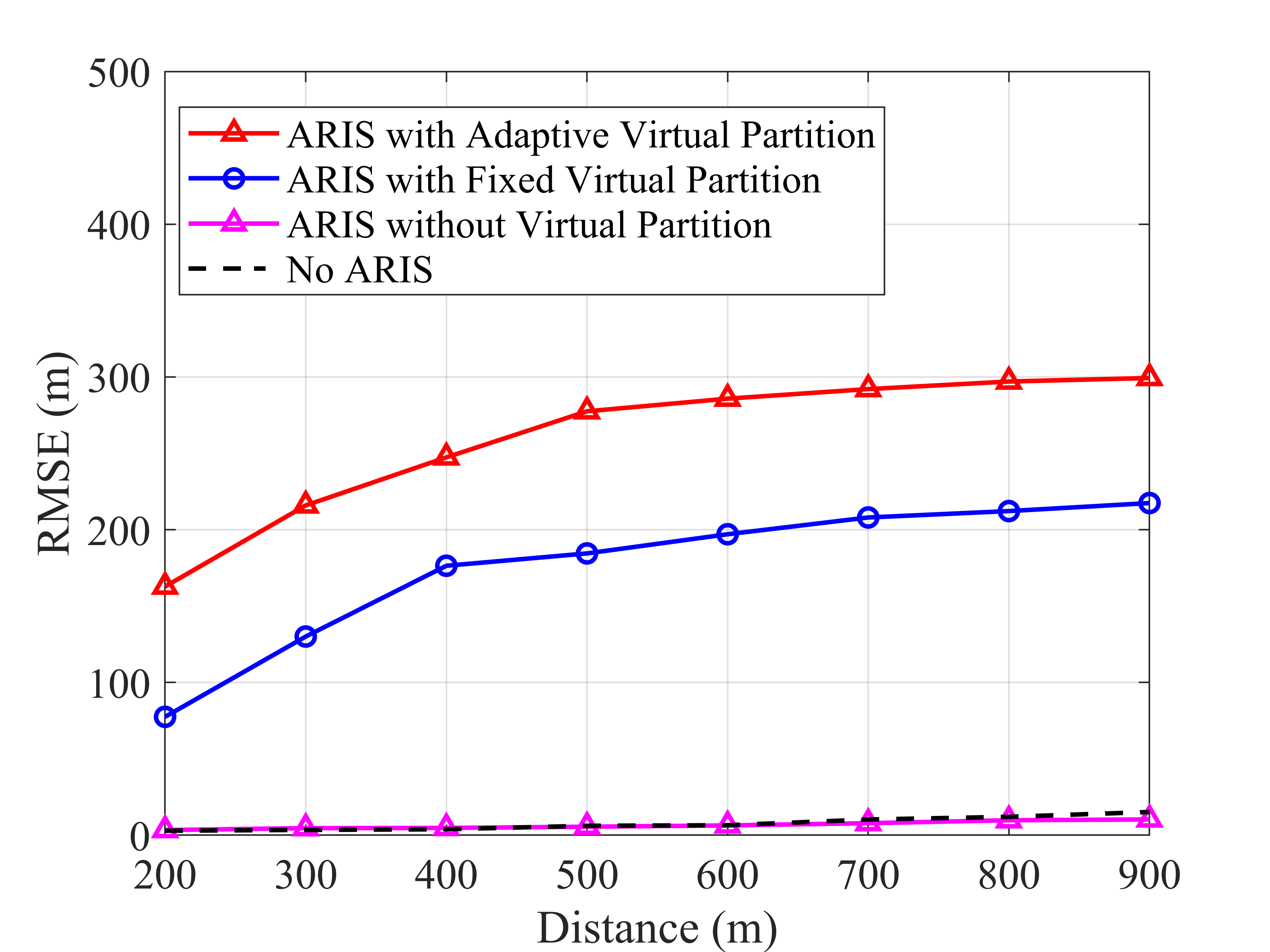} \label{fig3c}
	}
	\DeclareGraphicsExtensions.
	\caption{Simulation results for the sum-rate and the RMSE versus the distance between SU and MUs.}
	\label{dif}\vspace{-2mm}
\end{figure}
In this subsection, to observe the coverage performance of the proposed ARIS-based new architecture, we compare the sum-rate of RUs, and the RMSE of MU localization for the four schemes at different distances $d_{se}$. Additionally, the transmission power, and total ARIS power are set as $P_S^{\max} = 10$mW and $P_R^{\max} = 20$mW, respectively. The numbers of MUs and ARIS elements are set as $E=6$ and $N_t = 512$, respectively. Then, the weight $\omega$ is set as $\omega=0.15$.

As shown in Fig. \ref{dif}, when the distance between the MU and SU is large, i.e., $d_{se} > 400$m, the proposed scheme improves the RMSE by approximately 37.65\% while only sacrificing 3.69\% of the sum-rate, compared to the scheme with a fixed virtual partition and power allocation. When the distance is smaller, for example, when $d_{se} = 200$m, the proposed scheme reduces the rate by approximately 21.21\% and improves the RMSE by about 148.53\%, compared to the scheme with a fixed virtual partition and power allocation. According to (\ref{pal}), with fixed transmission power $P_S$, total ARIS power $P_R^{\max}$, and the number of MUs, the power allocation of ARIS mainly depends on the ratio of the distance from each MU to the ARIS and SU. Therefore, when the distance between the MUs and SU is small, the distance ratio increases, causing ARIS to allocate more power to enhance interference to the MUs.

These results indicate that the proposed scheme can adaptively adjust the power distribution based on the distance between the MU and SU, thereby ensuring effective interference to the MU without excessively sacrificing the sum rate.
\vspace{-2mm}
\subsection{Impact of the Total Power Limit}\label{effe}
\begin{figure}[!t]
	\centering
	\subfigure[Sum-rate of different schemes]{
		\includegraphics[width=0.4\textwidth,height=0.3\textwidth]
		{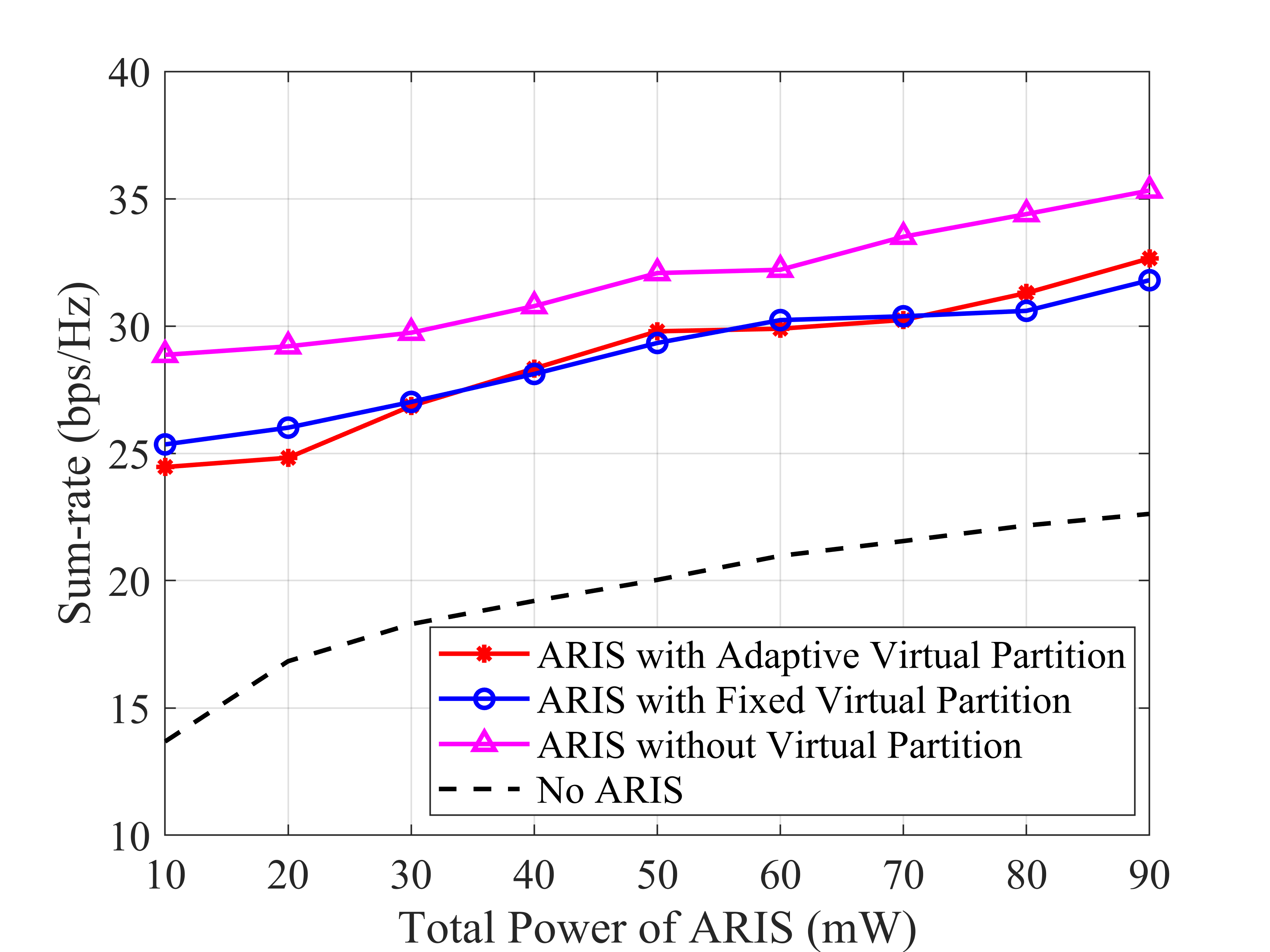}\label{fig5a}
	}
	\centering
	\subfigure[RMSE of different schemes]{
		\includegraphics[width=0.4\textwidth,height=0.3\textwidth]{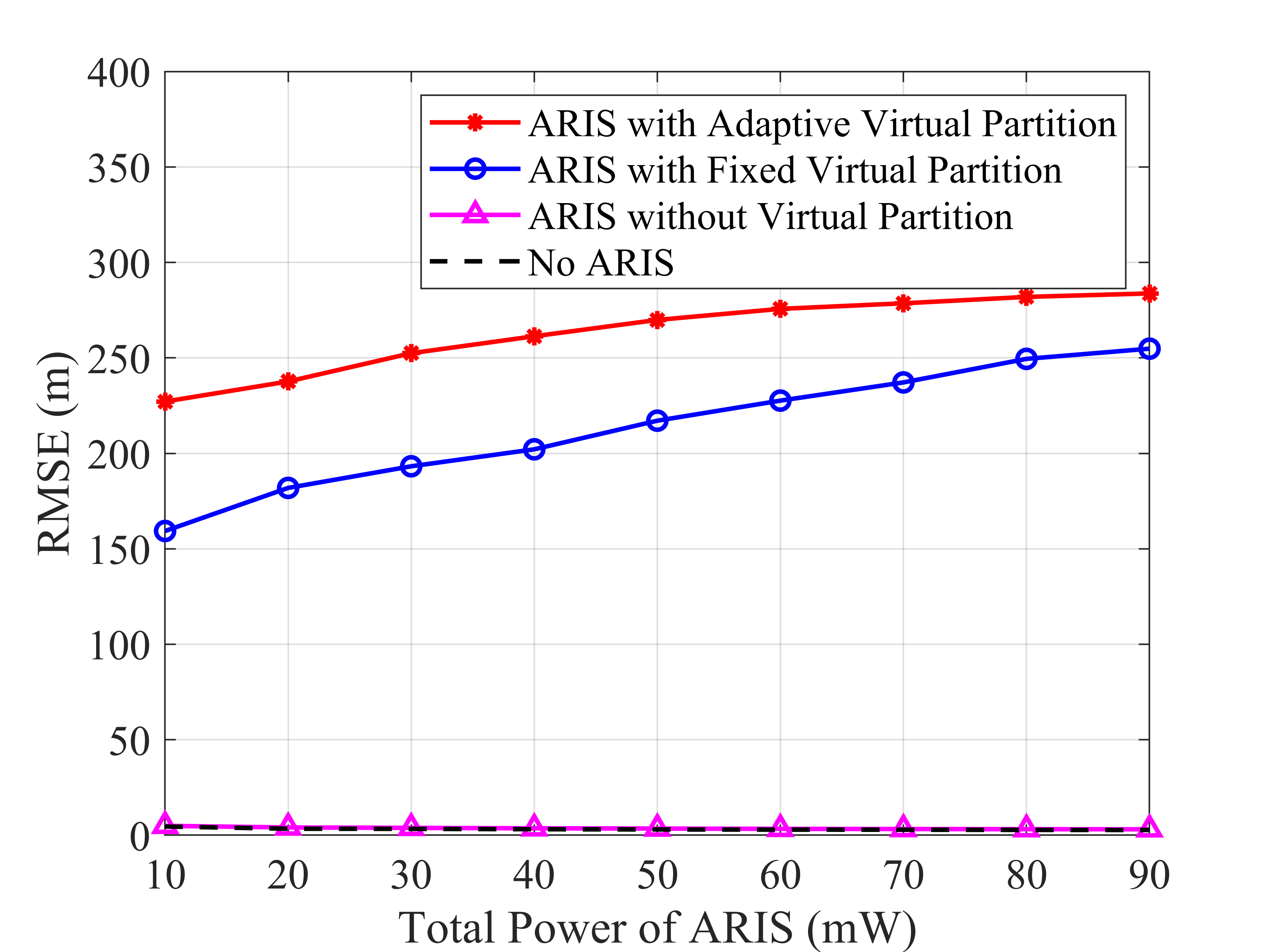} \label{fig5c}
	}
	\DeclareGraphicsExtensions.
	\caption{Simulation results for the sum-rate and the RMSE versus the total power $P_R^{max}$.}
	\label{P}
\end{figure}

In this subsection, to evaluate the impact of total power on the optimization performance of the proposed scheme, we compare the sum-rate of RUs, and the RMSE of MU localization for the four schemes at different total power of ARIS $P_R^{\max}$. Additionally, the transmission power is set as $P_S^{\max} = P_R^{\max} / 2$. The numbers of MUs and ARIS elements are set as $E=6$ and $N_t = 512$, respectively. Then, the weight $\omega$ is set as $\omega=0.15$ and the distance between MUs and SU is set as $d_{se} = 400$m.

As shown in Fig. \ref{P}, with the increase in the total power limit $P_R^{max}$, both the scheme without ARIS and the scheme of ARIS without virtual partition exhibit an upward trend in system sum rate, while the RMSE of MU localization decreases. This is because the increase in total power decreases the ISR at each MU, thereby enhancing its localization accuracy. In contrast, the proposed scheme not only increases the sum rate as the total power increases, but also improves the RMSE of MU localization. This is because, as the total power increases, ARIS can allocate more power to generate AN. Furthermore, compared to the scheme with a fixed virtual partition and power allocation, the proposed scheme improves the RMSE by approximately 36.51\% while only sacrificing 3.95\% of the sum-rate when the total power $P_R^{\max} \leq 20$mW. When 30mW $\leq P_R^{\max} \leq 70$mW, the proposed scheme improves the RMSE by approximately 24.20\% while maintaining a sum rate similar to that of the fixed scheme. Finally, it can be observed that the proposed scheme improves the sum rate by approximately 2.5\% and the RMSE by approximately 12.17\%. This is because the proposed scheme can effectively allocate ARIS elements and power for interfering with different MUs, thereby improving interference efficiency.
\vspace{-2mm}
\subsection{Impact of the Number of ARIS Reflection Elements}\label{effe2}
\begin{figure}[!t]
	\centering
	\subfigure[Sum-rate of different schemes]{
		\includegraphics[width=0.4\textwidth,height=0.3\textwidth]
		{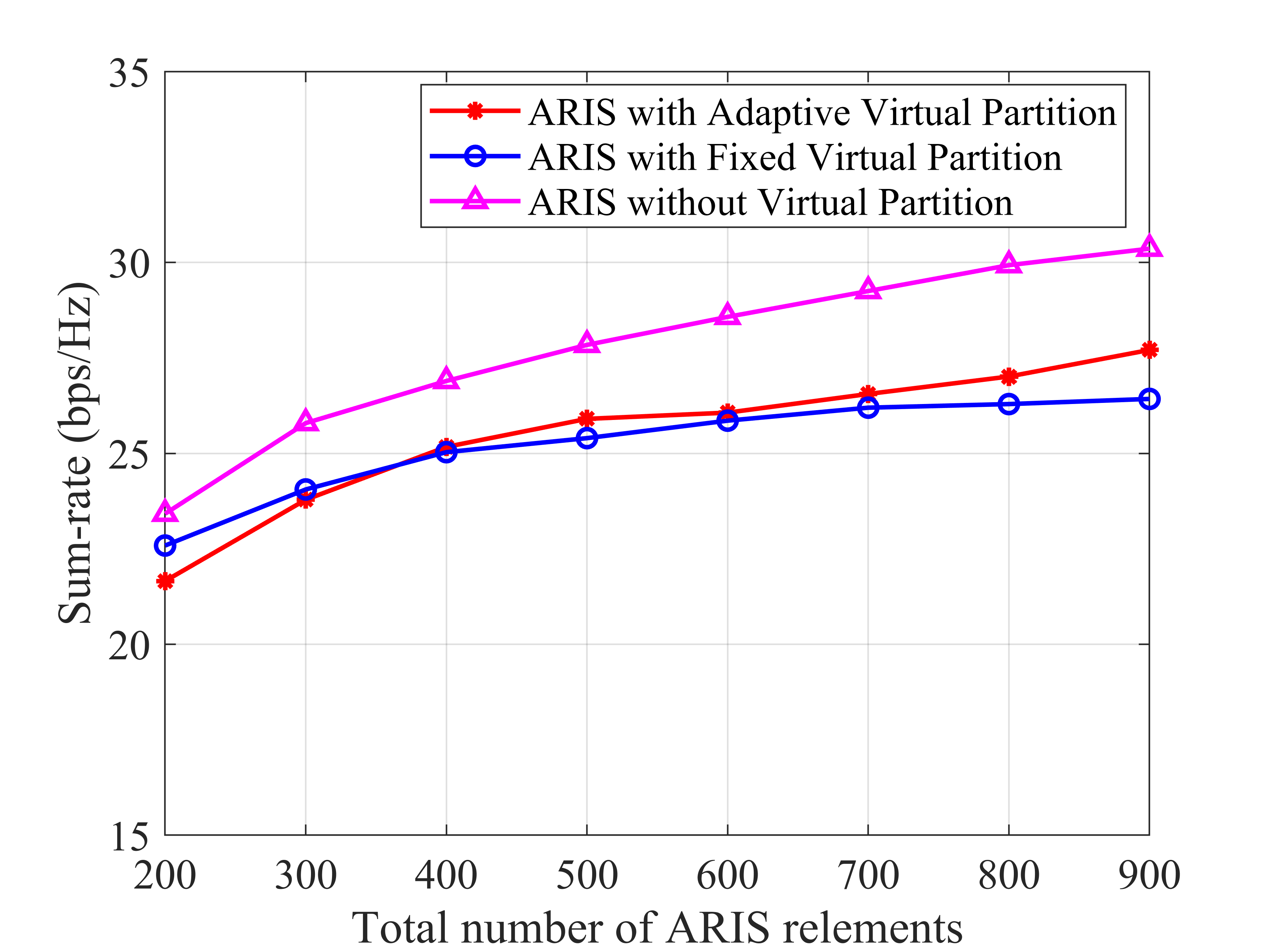}\label{fig6a}
	}
	\centering
	\subfigure[RMSE of different schemes]{
		\includegraphics[width=0.4\textwidth,height=0.3\textwidth]{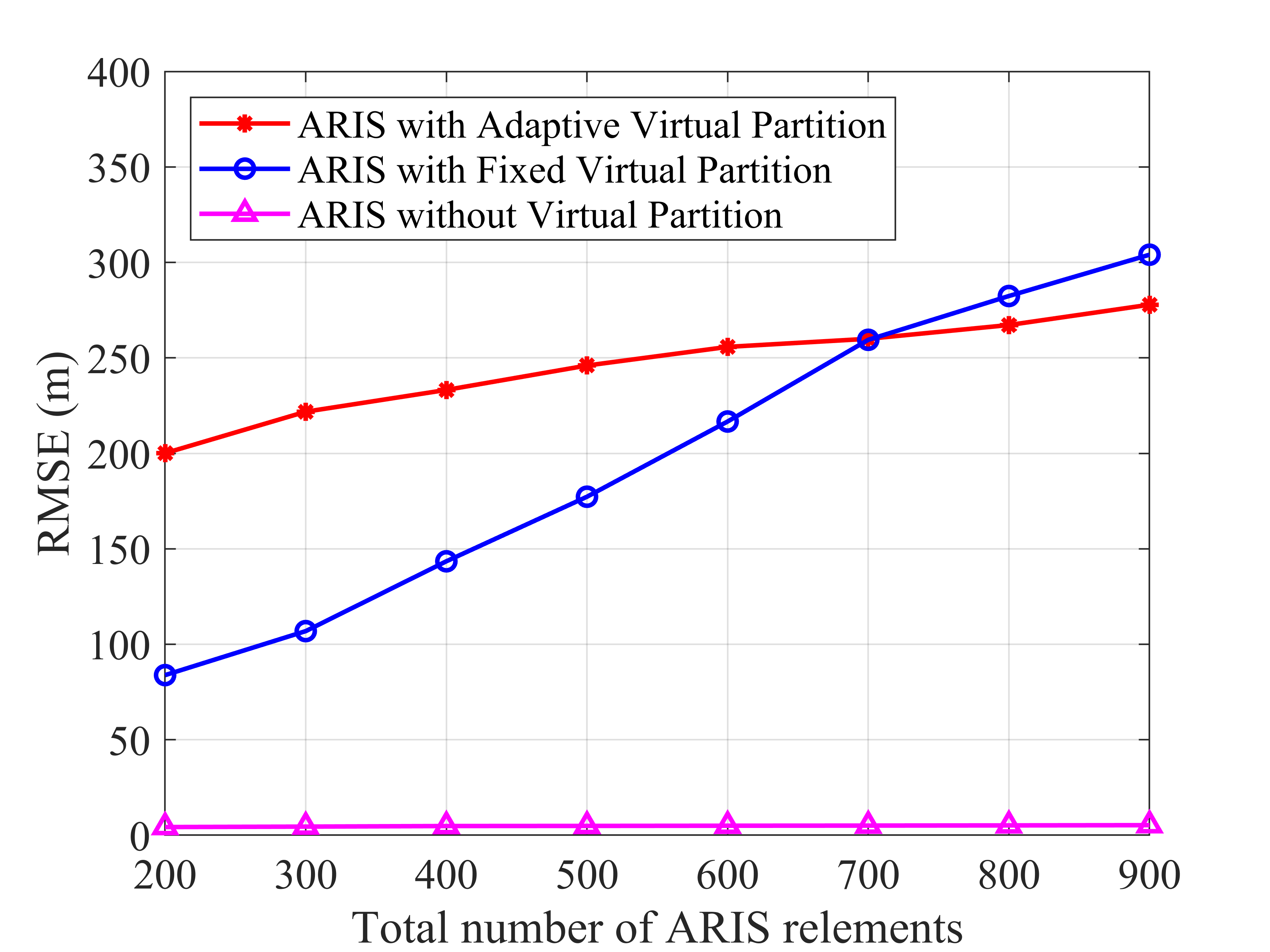} \label{fig6c}
	}
	\DeclareGraphicsExtensions.
	\caption{Simulation results for the sum rate and the RMSE versus the number of ARIS reflection elements $N_t$.}
	\label{RISU}
\end{figure}

In this subsection, to evaluate the impact of the ARIS elements on the optimization performance of the proposed scheme, we compare the sum rate of RUs and the RMSE of MU localization for the three schemes at different numbers of ARIS reflection elements $N_t$. Additionally, the transmission power and the total power of ARIS is set as $P_S^{\max} = 10$mW and $P_R^{\max} = 20$mW, respectively. The number of MUs is set as $E=6$. Then, the weight $\omega$ is set as $\omega=0.15$ and the distance between MUs and SU is set as $d_{se} = 400$m.

As shown in Fig. \ref{fig6a}, under limited ARIS reflection elements, the proposed scheme can significantly improve the MU localization RMSE while maintaining a rate similar to that of the fixed scheme. For example, when $N_t = 200$, the proposed scheme improves the MU localization RMSE by approximately 138.64\%. However, under sufficient ARIS reflection elements, the RMSE of the fixed scheme is slightly better than that of the proposed scheme, while the proposed scheme slightly outperforms the fixed scheme in terms of the sum rate. This is because, when ARIS reflection elements are limited, the proposed scheme can effectively partition the ARIS to improve interference efficiency while ensuring communication enhancement for the RUs. In contrast, when ARIS reflection elements are abundant, according to ($\ref{final}$), the proposed scheme adaptively allocates more elements to improve the sum rate while ensuring effective interference to the MUs.

\subsection{Impact of the Number of MUs}
\begin{figure}[!t]
	\centering
	\subfigure[Sum-rate of different schemes]{
		\includegraphics[width=0.4\textwidth,height=0.3\textwidth]
		{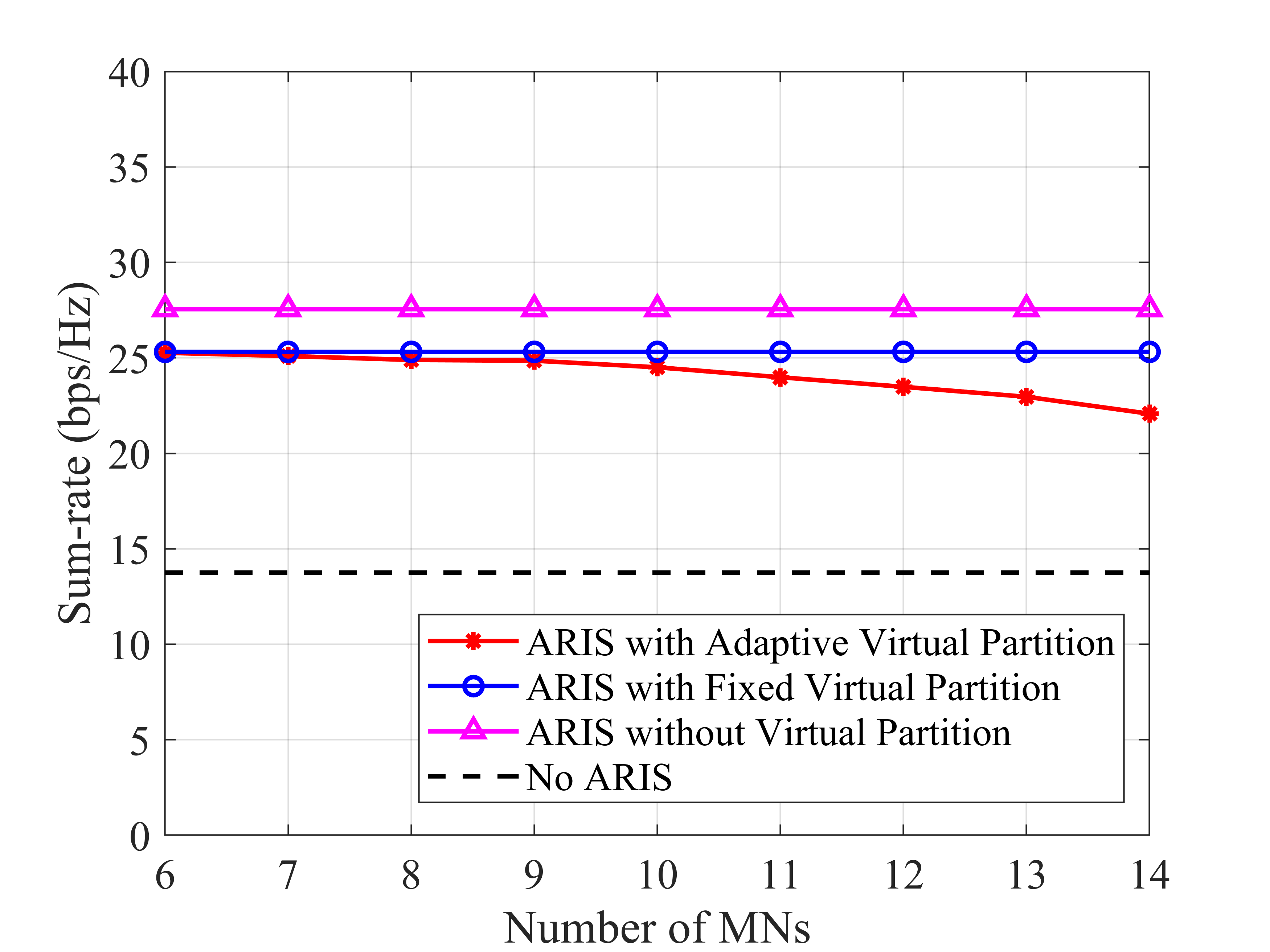}\label{fig7a}
	}
	\centering
	\subfigure[RMSE of different schemes]{
		\includegraphics[width=0.4\textwidth,height=0.3\textwidth]{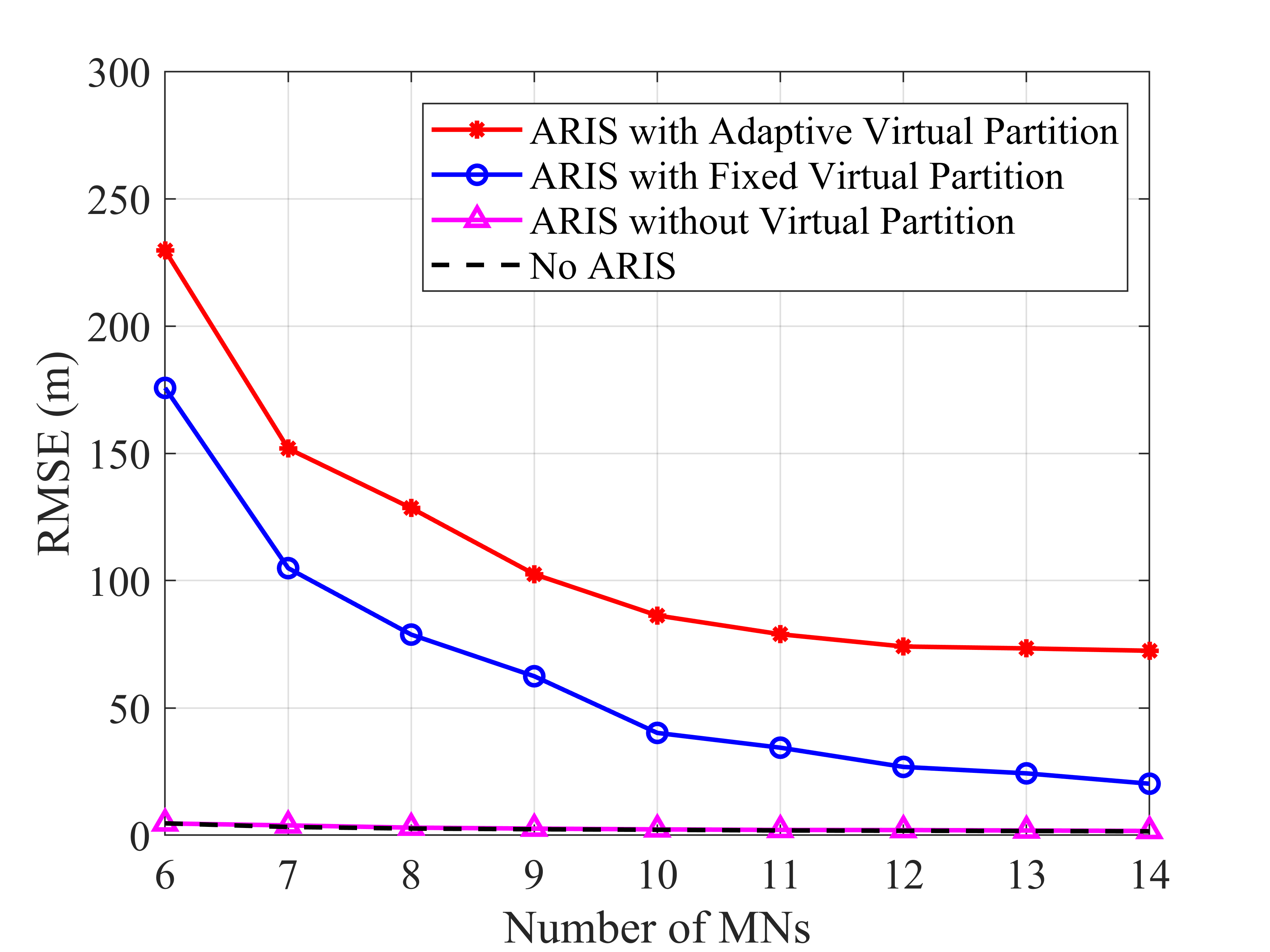} \label{fig7c}
	}
	\DeclareGraphicsExtensions.
	\caption{Simulation results for the sum rate and the RMSE versus the number of MNs $E$.}
	\label{RISE}
\end{figure}
In this subsection, to evaluate the impact of the number of MUs on the optimization performance of the proposed scheme, we compare the sum rate of RUs and the RMSE of MU localization for the four schemes at different numbers of MUs $E$. Additionally, the transmission power and the total power of ARIS is set as $P_S^{\max} = 10$mW and $P_R^{\max} = 20$mW, respectively.  The number of ARIS elements is set as $N_t = 512$, respectively. Then, the weight $\omega$ is set as $\omega=0.15$ and the distance between MUs and SU is set as $d_{se} = 400$m.

As shown in Fig. \ref{RISE}, as the number of MUs increases, the RMSE for all schemes decreases accordingly. Furthermore, the sum rate for the other schemes remains essentially unchanged while the sum rate for the proposed scheme decreases. When the number of MUs is 14, the proposed scheme improves the RMSE by 258.16\% at the cost of approximately 12.8\% in sum rate. This is because, as the number of MUs increases, the proposed scheme needs to allocate more ARIS elements and power to interfere with MU localization, thereby causing a decrease in its sum rate.
\vspace{-3mm}
\subsection{Impact of Optimization Weight $\omega$}\label{eff}
\begin{figure}[!t]
	\centering
	\includegraphics[width=3.2in]{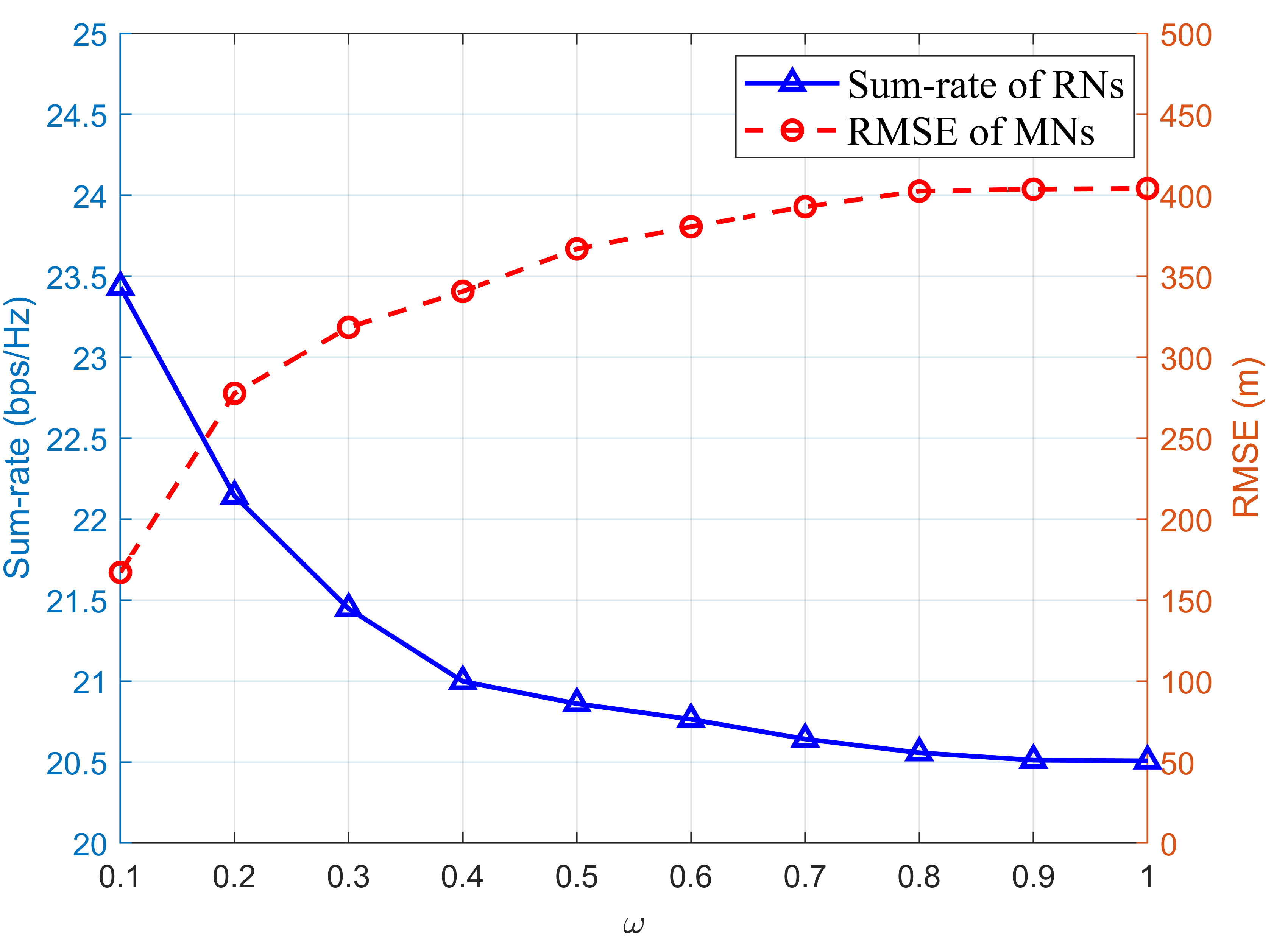}
	\caption{Performance of the proposed scheme with different weights $\omega$.}
	\label{lam}
\end{figure}
As a critical parameter in optimization problem (\ref{p}), the weight $\omega$ plays a decisive role in the final optimization result. Therefore, to explore the impact of different weights on performance, we plot the sum-rate of RUs, and the RMSE of MU localization at different weights $\omega$. Additionally, the transmission power and the total power of ARIS is set as $P_S^{\max} = 10$mW and $P_R^{\max} = 20$mW, respectively. The numbers of MUs and ARIS elements are set as $E=6$ and $N_t = 512$, respectively. Then, the distance between MUs and SU is set as $d_{se} = 400$m.

As shown in Fig. \ref{lam}, when $\omega<0.8$ increases, the sum rate of the proposed scheme decreases while its RMSE increases. When $\omega\geq0.8$, the sum rate and RMSE of the proposed scheme remain essentially unchanged. The reason for this phenomenon is that the weight $\omega$ primarily affects the number of ARIS-LI elements $N_e$. When $\omega<0.8$, as $\omega$ increases, ARIS allocates more reflection elements to interfere with the MUs. However, when $\omega\geq0.8$, ARIS needs to ensure communication enhancement for the RUs, at which point the number of ARIS-LI elements $N_e$ reaches its allocation limit.
%\vspace{-2mm}
\section{Conclusion}
\label{concl}
In this paper, we proposed a novel ARIS architecture integrated with an AN module to address the dual challenges of communication enhancement and location privacy protection in UAV networks. The architecture enhanced communication quality through intelligent control of reflected signals, while designing AN to interfere with attempts to locate the SU. Additionally, through adaptive ARIS partitioning, the joint optimization problem of communication enhancement and localization interference was decoupled in the physical space. Furthermore, we derived and validated the optimal solution for ARIS partitioning and power allocation under average channel conditions. Subsequently, by deriving the CRLB and formulating a multi-objective optimization problem, we designed efficient optimization algorithms for the reflection precoding and AN design in the ARIS-CE and ARIS-LI partitions, respectively. Simulation results validated that the proposed scheme could improve the legitimate communication rate and effectively protects the location privacy of the SU. Specifically, compared to the baseline scheme, the proposed solution improved the localization error by approximately 37.65\% with only a 3.69\% reduction in the sum rate. Additionally, the proposed scheme maintained good localization interference performance even under large-scale MUs.
\bibliography{reference}

@article{zhou2024delay,
	title={Delay-Aware UAV Computation Offloading and Communication Assistance for Post-Disaster Rescue},
	author={Zhou, Chengyi and Liu, Junyu and Qu, Kaige and Sheng, Min and Li, Jiandong and Zhuang, Weihua},
	journal={IEEE Trans. Wireless Commun.},
	volume={23},
	number={12},
	pages={19110--19125},
	month={Oct.},
	year={2024}
}

@article{sheng2025towards,
	title={Towards Disaster-Resistant Cellular Communication Networks Based on Network Capacity Scalability},
	author={Sheng, Min and Chen, Xuhui and Liu, Junyu and Li, Jiandong and Quek, Tony QS},
	journal={IEEE Trans. Wireless Commun.},
	volume={24},
	number={6},
	pages={5310--5322},
	month={Mar.},
	year={2025}
}

@article{guo2024joint,
	title={Joint Uplink and Downlink NOMA for UAV Relaying Network with Multi-Pair Users},
	author={Guo, Xu and Li, Bing and Wu, Jianjun and Zhang, Rongqing and Cheng, Xiang},
	journal={IEEE Trans. Wireless Commun.},
	volume={23},
	number={12},
	pages={18549--18562},
	month={Oct.},
	year={2024}
}

@article{lin2024multi,
	title={Multi-Antenna covert communication assisted by UAV-RIS with imperfect CSI},
	author={Lin, Shengbin and Xu, Yitao and Wang, Haichao and Ding, Guoru},
	journal={IEEE Trans. Wireless Commun.},
	volume={23},
	number={10},
	pages={13841--13855},
	month={Jun.},
	year={2024}
}

@article{yu2024secure,
	title={Secure ultra-reliable and low latency communication in {UAV}-enabled {NOMA} wireless networks},
	author={Yu, Kan and Feng, Zhiyong and Yu, Jiguo and Chen, Ting and Peng, Jinlin and Li, Dong},
	journal={IEEE Trans. Veh. Technol.},
	volume={73},
	number={10},
	pages={14908--14922},
	month={Jun.},
	year={2024}
}

@article{qu2024privacy,
	title={Privacy and security in ubiquitous integrated sensing and communication: Threats, challenges and future directions},
	author={Qu, Kaiqian and Ye, Jia and Li, Xuran and Guo, Shuaishuai},
	journal={IEEE Internet Things Mag.},
	volume={7},
	number={4},
	pages={52--58},
	month={Jun.},
	year={2024}
}

@article{gu2022physical,
	title={Physical layer security for {RIS}-aided wireless communications with uncertain eavesdropper distributions},
	author={Gu, Xiaohui and Duan, Wei and Zhang, Guoan and Sun, Qiang and Wen, Miaowen and Ho, Pin-Han},
	journal={IEEE Syst. J.},
	volume={17},
	number={1},
	pages={848--859},
	month={Mar.},
	year={2022}
}

@article{tong2024uav,
	title={{UAV}-assisted covert federated learning over mmWave massive {MIMO}},
	author={Tong, Ziheng and Wang, Jingjing and Hou, Xiangwang and Jiang, Chunxiao and Liu, Jianwei},
	journal={IEEE Trans. Wireless Commun.},
	volume={23},
	number={9},
	pages={11785--11798},
	month={Apr.},
	year={2024}
}

@article{li2024active,
	title={Active {RIS}-aided {NOMA}-enabled space-air-ground integrated networks with cognitive radio},
	author={Li, Junjie and Yang, Liang and Wu, Qingqing and Lei, Xianfu and Zhou, Fuhui and Shu, Feng and Mu, Xidong and Liu, Yuanwei and Fan, Pingzhi},
	journal={IEEE J. Sel. Areas Commun.},
	volume={43},
	number={1},
	pages={314--333},
	month={Sep.},
	year={2024}
}

@article{truong2025energy,
	title={Energy Efficiency in {RSMA}-Enhanced Active {RIS}-Aided Quantized Downlink Systems},
	author={Truong, Thanh Phung and Nguyen, Thi My Tuyen and Nguyen, The Vi and Dao, Nhu-Ngoc and Cho, Sungrae},
	journal={IEEE J. Sel. Areas Commun.},
	volume={43},
	number={3},
	pages={834--850},
	month={Jan.},
	year={2025}
}

@article{wen2024ris,
	title={{RIS}-assisted {UAV} secure communications with artificial noise-aware trajectory design against multiple colluding curious users},
	author={Wen, Yun and Chen, Gaojie and Fang, Sisai and Wen, Miaowen and Tomasin, Stefano and Di Renzo, Marco},
	journal={IEEE Trans. Inf. Forensics Security.},
	volume={19},
	pages={3064--3076},
	month={Jan.},
	year={2024}
}

@article{su2024secure,
	title={Secure Transmission Optimization for {RIS}-Aided {DFRC} Systems with Artificial Noise},
	author={Su, Ying and Dai, Zhutao and Peng, Zhangjie and Weng, Ruisong and Ren, Hong and Pan, Cunhua},
	journal={IEEE Commun. Lett.},
	volume={28},
	number={8},
	pages={1780--1784},
	month={Jun.},
	year={2024}
}

@article{chen2025double,
	title={Double-{RIS} Enabled Physical Layer Security for Wireless-Powered Communication Systems Over Rayleigh Fading Channels},
	author={Chen, Jingyu and Cao, Kunrui and Ding, Haiyang and Lv, Lu and Ye, Yinghui and Chi, Haolian and Wang, Tao and Yang, Liang},
	journal={IEEE Trans. Commun. (Early Access)},
	year={2025},
	publisher={IEEE}
}

@article{wang2024active,
	title={Active aerial reconfigurable intelligent surface assisted secure communications: Integrating sensing and positioning},
	author={Wang, Dawei and Wang, Zijun and Yu, Keping and Wei, Zhiqiang and Zhao, Hongbo and Al-Dhahir, Naofal and Guizani, Mohsen and Leung, Victor CM},
	journal={IEEE J. Sel. Areas Commun.},
	volume={42},
	number={10},
	pages={2769--2785},
	month={Jan.},
	year={2024}
}

@article{elsayed2025sum,
	title={Sum Secrecy Rate Optimization in {RIS}-Assisted {ISAC} Systems: A Manifold-Based Framework},
	author={Elsayed, Mohamed and Ibrahim, Ahmed S and Ismail, Mahmoud H and Samir, Ahmed},
	journal={IEEE Wireless Commun. Lett. (Early Access)},
	year={2025}
}

@article{han2022artificial,
	title={Artificial noise aided secure {NOMA} communications in {STAR-RIS} networks},
	author={Han, Yi and Li, Na and Liu, Yuanwei and Zhang, Tong and Tao, Xiaofeng},
	journal={IEEE Wireless Commun. Lett.},
	volume={11},
	number={6},
	pages={1191--1195},
	month={Mar.},
	year={2022}
}

@article{saif2025ris,
	title={{RIS} Alignment via Virtual Partitioning for Resilient Uplink Multi-{RIS}-Assisted {UAV} Communications},
	author={Saif, Mohammed and Valaee, Shahrokh},
	journal={IEEE Trans. Commun. (Early Access)},
	month={Jan.},
	year={2025}
}

@article{cai2023ris,
	title={{RIS} partitioning based scalable beamforming design for large-scale {MIMO}: Asymptotic analysis and optimization},
	author={Cai, Chang and Yuan, Xiaojun and Zhang, Ying-Jun Angela},
	journal={IEEE Trans. Wireless Commun.},
	volume={22},
	number={9},
	pages={6061--6077},
	month={Jan.},
	year={2023}
}

@article{sklar1997rayleigh,
	title={Rayleigh fading channels in mobile digital communication systems. I. Characterization},
	author={Sklar, Bernard},
	journal={IEEE Commun. Mag.},
	volume={35},
	number={7},
	pages={90--100},
	month={Jul.},
	year={1997}
}

@article{zhang2022active,
	title={Active {RIS} vs. passive {RIS}: Which will prevail in {6G}?},
	author={Zhang, Zijian and Dai, Linglong and Chen, Xibi and Liu, Changhao and Yang, Fan and Schober, Robert and Poor, H Vincent},
	journal={IEEE Trans. Commun.},
	volume={71},
	number={3},
	pages={1707--1725},
	month={Dec.},
	year={2022}
}

@article{lyu2023robust,
	title={Robust secure transmission for active RIS enabled symbiotic radio multicast communications},
	author={Lyu, Bin and Zhou, Chao and Gong, Shimin and Hoang, Dinh Thai and Liang, Ying-Chang},
	journal={IEEE Trans. Wireless Commun.},
	volume={22},
	number={12},
	pages={8766--8780},
	month={Apr.},
	year={2023}
}

@article{shen2018fractional,
	title={Fractional programming for communication systems—Part {I}: Power control and beamforming},
	author={Shen, Kaiming and Yu, Wei},
	journal={IEEE Trans. Signal Process.},
	volume={66},
	number={10},
	pages={2616--2630},
	month={Mar.},
	year={2018}
}

@article{boyd2011distributed,
	title={Distributed optimization and statistical learning via the alternating direction method of multipliers},
	author={Boyd, Stephen and Parikh, Neal and Chu, Eric and Peleato, Borja and Eckstein, Jonathan and others},
	journal={Found. Trends Mach. Learn.},
	volume={3},
	number={1},
	pages={1--122},
	month={Jul.},
	year={2011}
}

@article{shi2011iteratively,
	title={An iteratively weighted {MMSE} approach to distributed sum-utility maximization for a MIMO interfering broadcast channel},
	author={Shi, Qingjiang and Razaviyayn, Meisam and Luo, Zhi-Quan and He, Chen},
	journal={IEEE Trans. Signal Process.},
	volume={59},
	number={9},
	pages={4331--4340},
	month={Apr.},
	year={2011}
}

@article{hu2021two,
	title={Two-timescale channel estimation for reconfigurable intelligent surface aided wireless communications},
	author={Hu, Chen and Dai, Linglong and Han, Shuangfeng and Wang, Xiaoyun},
	journal={IEEE Trans. Commun.},
	volume={69},
	number={11},
	pages={7736--7747},
	month={Apr.},
	year={2021}
}

@article{abdallah2022ris,
	title={{RIS}-aided mmWave {MIMO} channel estimation using deep learning and compressive sensing},
	author={Abdallah, Asmaa and Celik, Abdulkadir and Mansour, Mohammad M and Eltawil, Ahmed M},
	journal={IEEE Trans. Wireless Commun.},
	volume={22},
	number={5},
	pages={3503--3521},
	month={Nov.},
	year={2022}
}

@article{wu2019towards,
	title={Towards smart and reconfigurable environment: Intelligent reflecting surface aided wireless network},
	author={Wu, Qingqing and Zhang, Rui},
	journal={IEEE Trans. Wireless Commun.},
	volume={58},
	number={1},
	pages={106--112},
	month={Nov.},
	year={2019}
}

@article{niu2022active,
	title={Active {RIS}-assisted secure transmission for cognitive satellite terrestrial networks},
	author={Niu, Hehao and Lin, Zhi and An, Kang and Liang, Xiaohu and Hu, Yihua and Li, Dong and Zheng, Gan},
	journal={IEEE Trans. Veh. Technol.},
	volume={72},
	number={2},
	pages={2609--2614},
	month={Sep.},
	year={2022}
}

@article{arzykulov2024aerial,
	title={Aerial {RIS}-aided physical layer security: Optimal deployment and partitioning},
	author={Arzykulov, Sultangali and Celik, Abdulkadir and Nauryzbayev, Galymzhan and Eltawil, Ahmed M},
	journal={IEEE Trans. Cogn. Commun. Netw.},
	volume={10},
	number={5},
	pages={1867--1882},
	month={Apr.},
	year={2024}
}

@article{martin2012modeling,
	title={Modeling and mitigating noise and nuisance parameters in received signal strength positioning},
	author={Martin, Richard K and King, Amanda Sue and Pennington, Jason R and Thomas, Ryan W and Lenahan, Russell and Lawyer, Cody},
	journal={IEEE Trans. Signal Process.},
	volume={60},
	number={10},
	pages={5451--5463},
	month={Jul.},
	year={2012}
}

@article{access2010further,
title={Further advancements for {E-UTRA} physical layer aspects (Release 9)},
journal={3GPP TS 36.814},
month={Mar.},
year={2010}
}

@article{mamaghani2022aerial,
	title={Aerial intelligent reflecting surface-enabled terahertz covert communications in beyond-5G Internet of Things},
	author={Mamaghani, Milad Tatar and Hong, Yi},
	journal={IEEE IEEE Internet Things J.},
	volume={9},
	number={19},
	pages={19012--19033},
	month={Mar.},
	year={2022}
}
\end{document}